\documentclass[aps, onecolumn,secnumarabic,amssymb,notitlepage,nobibnotes, nofootinbib,10pt]{revtex4-1}
\usepackage{graphicx}
\usepackage[pdftex,bookmarks,colorlinks,breaklinks]{hyperref}  
\usepackage{bm}
\usepackage{amsmath}
\setcitestyle{numbers}
\newcommand{\ber}{\begin{eqnarray}}
\newcommand{\ear}{\end{eqnarray}}
\newcommand{\bc}{\begin{center}}
\newcommand{\ec}{\end{center}}

\newcommand{\be}{\begin{equation}}
\newcommand{\ee}{\end{equation}}
\newcommand{\ba}{\begin{eqnarray}}
\newcommand{\ea}{\end{eqnarray}}
\newcommand{\bs}{\begin{subequations}}
\newcommand{\es}{\end{subequations}}

\newcommand{\forget}[1]{\iffalse#1\fi}
\newcommand{\forgetmenot}[1]{\iftrue#1\fi}

\begin{document}

\title{Post inflationary evolution of inflation-produced, large-scale magnetic fields using a generalised cosmological Ohm's law and both standard and modified Maxwell's equations}

\author{Timothy Oreta$^{1}$}
\author{Bob Osano$^{1,2}$}
\affiliation{$^{1}$Cosmology and Gravity Group, Department of Mathematics and Applied Mathematics, University of Cape Town ($UCT$), Rondebosch 7701, Cape Town, South Africa \\} 
\affiliation{$^{2}$ Centre for Higher Education Development ($CHED$), University of Cape Town ($UCT$), Rondebosch 7701, Cape Town, South Africa }

\begin{abstract}

In most literature on the evolution of cosmological magnetic fields, it is found that inflation-produced, large-scale magnetic fields evolved adiabatically. This rapid decay has been considered as the main obstacle against magnetic fields produced during the inflationary epoch from surviving until the present time and seeding the observed fields. However, recently, reports of the first-ever detection of intergalactic fields have emerged, with strengths around $10^{-6}G$ which are a mystery. One possible explanation is that magnetic fields could have been superadiabatically amplified in their evolutionary history. This can be demonstrated if we use the generalised cosmological Ohm's law and both standard and modified Maxwell equations; this is the goal of this study. We find that magnetic fields will be superadiabatically amplified when they are well over the horizon (after first horizon crossing) for the first time in their evolutionary history from the beginning of or during the epoch of radiation-domination ($RD$) or matter-radiation equality transition or matter-domination ($MD$) until much later during the $MD$ epoch when they cross the horizon for a second time and they go back to adiabatic decay until the present time. Hence, this will result in magnetic fields surviving until the present time and seeding the observed fields. This might be the explanation as to why magnetic fields with strengths of $10^{-6}G$ were detected.
 
\end{abstract}
\date{\today}
\maketitle

\section{\label{sec1} Introduction}

Although magnetic fields are pervasive in the Universe, none of the known magnetic generation mechanisms is able to account for the scope and breadth of what is observed. Several magnetic field generation mechanisms able to account for galactic magnetic fields have been devised and widely studied. These range from theories of primordial cosmological magnetic fields to astrophysical ones. The proposed mechanisms include magneto-genesis during the inflationary epoch \cite{And1,Tim29} and magnetic amplification. Magneto-genesis is the generation of a magnetic field in the early Universe. In this study we assume that magnetic fields are generated during the inflationary epoch. Magnetic amplification may mean increase in strength of the magnetic field or slower magnetic decay rates than the standard magnetic decay rate. However, these proposed mechanisms are not without challenges as will be discussed in this paper.

We examine the evolution of cosmological magnetic fields generated during inflation; regardless of the mechanism of generation. It is noteworthy that gravitational waves are also generated during inflation \cite{Kuro,bob1} as a counterpart. The energy transfer between the two is of interest and has been studied in \cite{Kuro}. In this present study we neglect counterpart gravitational waves.  In principle, inflation can generate magnetic fields on all scales although non-standard physics may have to be invoked to achieve non-minimal coupling of the electromagnetic field \cite{Tim30} to either matter or spatial curvature. In some cases, one may need a mechanism for breaking the conformal-invariance \cite{Tim31}. In order to examine if there is a link between the magnetic fields produced during inflation and those observed in the present Universe, it is crucial to understand how cosmological magnetic fields evolved during the epochs of $RD$, matter-radiation equality transition (assuming that the transit is an epoch and the reasons to that assumption will be given later in this paper) and $MD$; while assuming a spatially flat Friedmann Lemaitre Robertson Walker ($FLRW$) Universe.

The evolution of cosmological magnetic fields embedded in an expanding Universe whose expansion is variously dominated by radiation, matter and dark energy is still an open question. But the expansion of the Universe is usually broken into different intervals, namely inflation (for magneto-genesis and sometimes evolution of magnetic fields), $RD$, $MD$ and dark-energy dominated epochs. Magnetic fields are usually analyzed separately in these epochs and assumptions are made about their evolution during the transition period between epochs. The matter-radiation transition is of particular interest because it lasted sufficiently long, about 23000 years \cite{Tim46} which is much longer than the duration of the inflation or reheating epoch, thereby affording time for magnetic fields to change appreciably. Therefore, and additionally, studying the evolution of magnetic fields during the matter-radiation equality transit epoch enriches the standard picture of the evolutionary history of the magnetic fields. In analysing these epochs we employ the fluid approach because in order for us to probe the evolution of magnetic fields during the matter-radiation equality epoch we use the multi-fluid formalism which stems from using the single-fluid formalism in analyzing the evolution of magnetic fields first during the $RD$ epoch and then during the $MD$ epoch (before analyzing the evolution of magnetic fields during the $MD$ epoch using the single-fluid formalism, we analyze the evolution during the matter-radiation equality epoch first using the multi-fluid approach after analyzing the evolution during the $RD$ epoch using the single-fluid formalism).

In most of the literature on evolution of magnetic fields, it is found that magnetic fields evolved as $B^{2}\propto a^{-4}$ (adiabatic magnetic decay) for the flat Friedman model \cite{Tim32}. This rapid decay has been considered as the main obstacle against magnetic fields produced during the inflationary epoch from surviving until today and seeding the observed fields \cite{Tim32}. Conventional magnetic fields in spatially flat $FLRW$ Universes decay adiabatically throughout the evolution of these models and on all scales \cite{Tim32}. Inflation naturally achieves superhorizon correlations, hence it can easily produce primordial fields. Magnetic fields generated just after or way after inflation are too small in scale \cite{Tim32}. Magnetic fields produced during inflation decay adiabatically as soon as they cross outside the Hubble horizon. The result are astrophysically irrelevant magnetic fields in the present time \cite{Tim32}.
 
As already mentioned, recent detection of possible intergalactic fields, with strengths around $10^{-6} G$ demand an explanation. Superadiabatic amplification may be an explanation. Superadiabatic amplification may mean increase in strength of the magnetic field or slower magnetic decay rates than the standard (adiabatic) magnetic decay rate. It can be demonstrated if we use the generalised cosmological Ohm's law and both standard and modified Maxwell equations as we will show in this paper; this is the goal of this paper. But first let us consider modified Maxwell's equations for the reasons that will be given in the following paragraphs.
 
A homogeneous and isotropic Universe with a uniformly distributed net charge cannot be described by standard Maxwell's equations, because this requires that the electromagnetic field tensor in such a Universe must vanish everywhere \cite{Tim27}. Standard Maxwell's equations always fail for a closed Universe with a non-zero net charge regardless of the spacetime symmetry and the charge distribution \cite{Tim27}. Hence, modifications of systems of equations seem necessary. Such changes can be implemented via Maxwell's equations. In particular, non-trivial modifications of Maxwell's equations are required. A Proca-type equation which contains a photon mass term is a type of modified Maxwell's equations \cite{Tim27}. These electromagnetic field equations can naturally arise from spontaneous symmetry breaking or the Higgs mechanism in quantum field theory ($QFT$), where photons acquire a mass by devouring massless Goldstone bosons. However, when the symmetry is restored, photons loose their mass again and the problems mentioned above reappear. The second type of modification is where an electromagnetic field potential vector is coupled to the spacetime curvature tensor \cite{Tim27}. This type of electromagnetic field equations return to Maxwell's equations in a flat or Ricci-flat spacetime and don't introduce a new dimensional parameter. It was shown in \cite{Tim27} that modifications of Maxwell's equations have no discernible impact on existing experiments and observations. The impact of a curvature term coupled to field equations is in theory testable in astrophysical environments where the mass density is high or the gravity of electromagnetic radiation plays a dominant role in the dynamics of the system such as in the interior of neutron stars and during the early Universe. 

A photon will have an effective, time-dependent mass due to the additional terms $RA^{2}$ or $R_{ab}A^{a}A^{b}$ or $RA^{2}+R_{ab}A^{a}A^{b}$ \cite{And1} where $R$ is the curvature scalar, $R_{ab}$ is the Ricci tensor, $A^{a}$ is the four vector electromagnetic potential and $a$ and $b$ are spacetime indices. This is undesired as charge conservation is broken \cite{And1}. Nevertheless, expected effects (due to the additional terms) which contradict present-day observations or experiments are not observed \cite{And1}. Due to these terms we have $m_{\gamma}\sim R^{\frac{1}{2}}$ as the mass of the photon and $R^{\frac{1}{2}}\sim H$ , where $H$ is the expansion rate of the Universe \cite{And1}. Today the photon mass would be $m_{\gamma}\sim H_{today}\sim10^{-33}eV$ , which is well below the present limits to the photon mass, that is, $m_{\gamma}<3\times10^{-27}eV$ \cite{Tim25}. Though charge non-conservation would only manifest itself on scales of the horizon or larger, this has no observable consequences. The terms $RA^{2}+R_{ab}A^{a}A^{b}$, $RA^{2}$ and $R_{ab}A^{a}A^{b}$ vanish in vacuum and as a consequence they cannot affect the propagation of photons outside massive bodies. For a $RD$ Universe, these terms introduce corrections of order $\frac{H^{2}}{T^{2}}\sim\frac{T^{2}}{m_{pl}}$ where $T$ is temperature and $m_{pl}$ is the Planck mass. These corrections are negligible for temperatures where the evolution of the Universe is relatively well understood. Therefore, these terms cannot spoil successful predictions made using the standard Maxwell equations \cite{And1}. Hence, due to the brief explanation above on the terms, we will consider them. However, despite the need to modify standard Maxwell's equations, we will show that superadiabatic amplification of magnetic fields can be achieved by using the standard Maxwell's equations and the generalised cosmological Ohm's law.

As a result of all this discussion, it seems that the present-day amplitude of magnetic fields arising from inflationary magnetogenesis can actually be much larger than what has been claimed in most previous studies. We will show this in the sections to follow.

This paper is organised as follows: In Section $II$ we introduce the Faraday tensor and the Bianchi identity, we discuss the generalised cosmological Ohm's law, matter action, electromagnetic action, coupling action, the derivation of modified Maxwell tensors from the total action of the fluid action, the Maxwell action, the Coulomb action and the action due to the coupling of the terms $RA^{2}+R_{ab}A^{a}A^{b}$ or $RA^{2}$ or $R_{ab}A^{a}A^{b}$ using the single fluid formalism. In Section $III$ we discuss the action principle for a two-conducting fluid approximation and the derivation of the modified Maxwell tensors from the same total action mentioned before above except that for this case we are considering two fluids hence, we use the two (multi)-conducting fluid formalism. In Section $IV$ we use the modified Maxwell tensors derived using both the single and multi-conducting fluid formalisms to study the evolution of magnetic fields during the $RD$, matter-radiation equality and $MD$ epochs. We also discuss large-scale superadiabatic magnetic amplification during these epochs. In section $V$ we analyse the role of initial conditions in achieving superadiabatic amplification during the mentioned epochs before above. In section $VI$ we summarise our findings for the epochs of $RD$, matter-radiation equality and $MD$. We discuss and summarise everything in this paper in Section $V$. For the spacetime indices we have $a$, $b$=0, 1, 2, 3.

\section{\label{sec2} The action principle for a single-conducting fluid approximation and modified Maxwell equations}

In this section an action principle is set up to derive the modified Maxwell equations \cite{Leo}. The pull-back formalism will be used to set up variations of $A_{a}$ required to get the modified Maxwell equations \cite{Leo}. The modified Maxwell equations are obtained from a coupling term based on the scalar $J^{a}_{T}A_{a}$ where $J^{a}_{T}$ is the total current of the fluid composed of the flux current of the fluid and the plasma current from the generalised cosmological Ohm's law (since we are considering a single-conducting fluid). The other way is by varying $A_{a}$ , which will appear in two pieces of the total action: One constructed from the antisymmetric Faraday tensor $F_{ab}$, defined as \cite{Leo}
\begin{eqnarray}
F_{ab}&=&\nabla_{a}A_{b}-\nabla_{b}A_{a},
\end{eqnarray}
and satisfies a Bianchi identity
\begin{eqnarray}
\nabla_{a}F_{bc}+\nabla_{c}F_{ab}+\nabla_{b}F_{ca}&=&0.
\end{eqnarray}
It is important to note that the formalism will account for coupling of a fluid to dynamical spacetime \cite{Leo}. The fluid action $S_{M}$ will have as its Lagrangian an energy functional $\Lambda$ \cite{Leo} (also known as the Master function). In the following subsection, we will briefly discuss the generalised cosmological Ohm's law before we consider a system with a single-fluid flow where the generalised cosmological Ohm's law is used.

\subsection{Discussion on the generalised cosmological Ohm's law}

One typically applies Ohm's law in order to solve for the evolution of the electromagnetic fields and the plasma \cite{Tim26}. Given, $\vec{J}_{q}=\sigma(\vec{E}+\vec{v}\times\vec{B})$, where $\vec{v}$ is the bulk velocity of the plasma, $\sigma$ is the conductivity of plasma that has units of an inverse length, $\vec{J}_{q}$ is current due to plasma, $\vec{E}$ is the electric field and $\vec{B}$ is the magnetic field, Ohm's law represents a shortcut to solving both standard or modified Maxwell's equations by providing a link between current and the electromagnetic fields \cite{Tim26}. After assessing the relevant evolution and interaction timescales, Ohm's law derives from the evolution equation of the current and reduces to the simple form before above \cite{Tim26}. It is also possible to derive a generalised relativistic Ohm's law and the current evolution although we will not discuss it in this paper. For the derivation of both the generalised relativistic Ohm's law and the current evolution see (\cite{Tim41} and references therein).
 
We now give a brief overview on the derivation of the generalised Ohm's law relevant on cosmological scales (for a detailed discussion check \cite{Tim26}). Linearising equation (2.25) in the article \cite{Tim26} for a $FLRW$ background will not alter its main features. As a result of working in an arbitrary frame with four velocity, $u^{a}$ in the presence of the proton-electron centre of mass current, $J^{a}_{pe}$ we have $J^{a}_{pe}=0$. Again, $J^{a}_{pe}=0$ if we choose $u^{a}$ to be the centre of mass frame, which is very close to the baryon frame. We express the spatially projected 4-vectors with respect to a comoving basis, i.e. $J^{a}=a^{-1}(0, \vec{J})$ where $a$ is the scale factor, and $\vec{J}$ is the 3-current due to plasma. Then we find the following linear current evolution equation
\begin{eqnarray}
\dot{\vec{J}}+(4H+\Gamma_{C}+\Gamma_{T})\vec{J}&=&-\Gamma_{T}\vec{J}_{r}+\omega^{2}_{p}\vec{E}
\end{eqnarray}
where $H$ is the Hubble parameter which comes from $\Theta=3H+\mathcal{O}(1)$, $\Gamma_{C}$ is the Coulomb rate, $\Gamma_{T}$ is the Thomson rate, $\omega_{p}$ is the plasma frequency, $\vec{J}_{r}$ is the photon current and $\vec{E}$ is the electric field all elaborated in \cite{Tim26}. Approximating the time derivative of the current by a characteristic timescale of the problem, $\tau$, through $\dot{\vec{J}}\simeq\tau^{-1}\vec{J}$, we can write a generalised cosmological Ohm's law in a familiar form. $\tau\sim$ min$(L, H^{-1})$ typically for large-scale fluctuations of physical correlation lengths $L$ larger than the Silk damping scale. Therefore, the linear cosmological Ohm's law can be written as
\begin{eqnarray}
(\eta_{\tau}+4\eta_{H}+\eta_{C}+\eta_{T})\vec{J}+\eta_{T}\vec{J}_{r}\simeq\vec{E}
\end{eqnarray}
where the different resistivities are
\begin{eqnarray}
\eta_{\tau}\equiv\frac{\tau^{-1}}{\omega^{2}_{p}} , \nonumber\ \eta_{H}\equiv\frac{H}{\omega^{2}_{p}} , \nonumber\ \eta_{C}\equiv\frac{\Gamma_{C}}{\omega^{2}_{p}} , \nonumber\ \eta_{T}\equiv\frac{\Gamma_{T}}{\omega^{2}_{p}}
\end{eqnarray}
and have the dimensions of time. With the presence of an electric field the resistivities quantify the efficiency of generating currents. You will notice that $\eta_{T}\gg\eta_{C}$ for $a\lesssim3\times10^{-6}$, well before recombination, and then $\eta_{T}\ll\eta_{C}$ until the present time when the resistivities are compared as functions of the scale factor in the upper panel of figure $1$ in the article \cite{Tim26}. $\eta_{H}$ which is resistivity due to expansion is small. It is proportional to the Hubble parameter and overcome by much faster interaction rates. $\eta_{\tau}$ which is the characteristic evolution timescale is negligible since it is inversely proportional to the large scales we consider. This means that the time derivative in the current evolution equation can safely be neglected. This implies that equation (4) can be written in the form
\begin{eqnarray}
\vec{J}+\sigma_{E}\eta_{T}\vec{J}_{r}\simeq\sigma_{E}\vec{E}
\end{eqnarray}
where $\sigma_{E}\equiv(\eta_{C}+\eta_{T})^{-1}$ is the electric conductivity of the plasma with the dimensions of an inverse length. 

With the above discussion, we can now consider a system with a single-fluid flow in the following subsection. We will discuss the matter, electromagnetic and coupling actions first.

\subsection{The matter, electromagnetic, and coupling actions}

In order to examine the effect of fluid couplings one needs a formalism that explicitly expresses the coupling in terms of the different fluid parameters. The most suitable formalism for this is the convective variational formalism \cite{Carter, And5,Tim12, Tim8, BT}. In this formalism, it can be shown that the fluid action $S_{M}$ has as Lagrangian the Master function $\Lambda$, which depends on the $n^{2}_{X}=-n^{X}_{a}n^{a}_{X}$, where $n^{a}_{X}$ is the number density four current or flux whose magnitude $n_{X}$ is the particle number density (see \cite{And5}), $X$ is a label for a fluid and $g_{ab}$ is the metric. Ignoring boundary terms throughout, an arbitrary variation of $S_{M}$ with respect to the flux $n^{a}_{X}$ and the metric results in (see\cite{Leo})
\begin{eqnarray}
\delta S_{M}&=&\delta\Bigg(\int_{\mathcal{M}}d^{4}x\sqrt{-g}\Lambda\Bigg)\nonumber=\int_{\mathcal{M}}d^{4}x\sqrt{-g}\Bigg[\mu^{X}_{a}\delta n^{a}_{X}+\frac{1}{2}(\Lambda g^{ab}+n^{a}_{X}\mu^{b}_{X})\delta g_{ab}\Bigg],
\end{eqnarray}
where $\mathcal{M}$ is the manifold or hypersurface, $g$ is the determinant of the metric, and $\mu^{X}_{a}$ are the canonically conjugate momenta to $n^{a}_{X}$; that is, letting \cite{Leo}
\begin{eqnarray}
\mathcal{B}^{X}&=&-2\frac{\partial\Lambda}{\partial n^{2}_{X}},
\end{eqnarray}
then
\begin{eqnarray}
\mu^{X}_{a}&=&g_{ab}\mathcal{B}^{X}n^{b}_{X}
\end{eqnarray}
(see \cite{Leo}).

 The Maxwell action is given by \cite{Leo} 
\begin{eqnarray}
S_{Max}&=&\frac{1}{16\pi}\int_{\mathcal{M}}d^{4}x\sqrt{-g}F_{ab}F^{ab},
\end{eqnarray}
and varying this action with respect to the four-vector potential $A_{a}$ (this couples the charged fluids to the electromagnetic field and vice versa) and the metric $g_{ab}$ results in \cite{Leo}
\begin{eqnarray}
\delta S_{Max}&=&\frac{1}{4\pi}\int_{\mathcal{M}}d^{4}x\sqrt{-g}(\nabla_{a}F^{ab})\delta A_{b}-\frac{1}{32\pi}\int_{\mathcal{M}}d^{4}x\sqrt{-g}(F_{cd}F^{cd}g^{ab}-4F^{ac}F^{b}_{c})\delta g_{ab}.
\end{eqnarray}
Since we are considering a single-conducting fluid, the minimal coupling of the Maxwell field to the charge current densities is obtained from
\begin{eqnarray}
S_{C}&=&\int_{\mathcal{M}}d^{4}x\sqrt{-g}(J^{a}_{X}+j^{a})A_{a},
\end{eqnarray}
where $J^{a}_{X}=e_{X}n^{a}_{X}$ is flux current, $j^{a}$ is four current due to plasma which can be rewritten as $j^{a}=(j^{0}, \vec{J})=(j^{0}, \vec{j})$. Here, $j^{0}=c\rho$ is the time part, $c$ is the speed of light, $\rho$ is the charge density of plasma and $\vec{J}=\vec{j}$ is the 3-current due to plasma which will be represented by the generalised cosmological Ohm's law in simplified form or equation (5) later on in this paper. Varying equation (10) with respect to $n^{a}_{X}$, $A_{a}$, $g_{ab}$, $J^{a}_{X}$ and $j^{a}$ results in
\begin{eqnarray}
\delta S_{C}&=&\int_{\mathcal{M}}d^{4}x\sqrt{-g}\Bigg[(J^{a}_{X}+j^{a})\delta A_{a}+e_{X}A_{a}\delta n^{a}_{X}+A_{a}\delta j^{a}+\frac{1}{2}(J^{a}_{X}+j^{a})A_{a}g^{bc}\delta g_{bc}\Bigg].
\end{eqnarray}
For a given system, the total action will thus become
\begin{eqnarray}
\delta S&=&\delta S_{M}+\delta S_{Max}+\delta S_{C}\nonumber\\
&=&\int_{\mathcal{M}}d^{x}\sqrt{-g}\Bigg\{[\mu^{X}_{a}+e_{X}A_{a}]\delta n^{a}_{X}+\frac{1}{4\pi}[\nabla_{b}F^{ba}+4\pi( J^{a}_{X}+j^{a})]\delta A_{a}+\frac{1}{2}\Bigg[\Lambda g^{ab}\nonumber\\
&+&n^{a}_{X}\mu^{b}_{X}+(J^{c}_{X}+j^{c})A_{c}g^{ab}-\frac{1}{16\pi}(F_{cd}F^{cd}g^{ab}-4F^{ac}F^{b}_{c})\Bigg]\delta g_{ab}+A_{a}\delta j^{a}\Bigg\}
\end{eqnarray}
where $J^{a}_{X}+j^{a}=J^{a}_{T}$. Now in the following subsections and sections, we will consider the case where the terms $RA^{2}+R_{ab}A^{a}A^{b}$, $RA^{2}$ and $R_{ab}A^{a}A^{b}$ are added to the action (12) before above. We first do so in the following subsection.

\subsection{The field equations}

 We now consider
\begin{eqnarray}
S_{\phi_{T}}&=&S_{\phi}+S_{\phi_{0}}
\end{eqnarray}
where $S_{\phi T}$ is the total action with coupling constants $\phi$ and $\phi_{0}$ in $\phi RA^{2}+\phi_{0} R_{ab}A^{a}A^{b}$ .
Varying action (13) yields
\begin{eqnarray}
\delta S_{\phi_{T}}&=&\delta S_{\phi}+\delta S_{\phi_{0}}
\end{eqnarray}
and writing equation (14) explicitly results in,
\begin{eqnarray}
\delta S_{\phi_{T}}&=&\int_{\mathcal{M}}d^{4}x\sqrt{-g}(2\phi RA^{a}+2\phi_{0} R^{a}_{b}A^{b})\delta A_{a}-\int_{\mathcal{M}}d^{4}x\sqrt{-g}\phi g^{ha}g^{fb}R_{fh}A^{2}\delta g_{ab}\nonumber\\&+&2\int_{\mathcal{M}}d^{4}x\sqrt{-g}\phi g^{ab}A^{2}\Bigg\{\frac{1}{2}\nabla_{c}[g^{cd}(\nabla_{a}\delta g_{db}+\nabla_{b}\delta g_{ad}-\nabla_{d}\delta g_{ba})]\nonumber\\&+&\frac{1}{2}g_{fh}\nabla_{b}\nabla_{a}\delta g^{fh}\Bigg\}+\frac{1}{2}\int_{\mathcal{M}}d^{4}x\sqrt{-g}g^{ab}(\phi RA^{2}+\phi_{0}R_{cd}A^{c}A^{d})\delta g_{ab}.
\end{eqnarray}
The total action then takes the form
\begin{eqnarray}
\delta S_{T}&=&\delta S_{M}+\delta S_{Max}+\delta S_{C}+\delta S_{\phi_{T}}
\end{eqnarray}
and writing explicitly,
\begin{eqnarray}
\delta S_{T}&=&\int_{\mathcal{M}}d^{4}x\sqrt{-g}\Bigg\{[\mu^{X}_{a}+e_{X}A_{a}]\delta n^{a}_{X}+\frac{1}{4\pi}[\nabla_{b}F^{ba}+4\pi(J^{a}_{X}+j^{a})+8\pi\phi RA^{a}+8\pi\phi_{0} R^{a}_{b}A^{b}]\delta A_{a}\nonumber\\
&+&\frac{1}{2}\Bigg[\Lambda g^{ab}+n^{a}_{X}\mu^{b}_{X}+J^{c}_{X}A_{c}g^{ab})-\frac{1}{16\pi}(F_{cd}F^{cd}g^{ab}-4F^{ac}F^{b}_{c})-\phi g^{ha}g^{fb}R_{fh}A^{2}\Bigg]\delta g_{ab}\nonumber\\
&+&2\phi g^{ab}A^{2}\Bigg[\frac{1}{2}\nabla_{c}(g^{cd}(\nabla_{a}\delta g_{db}+\nabla_{b}\delta g_{ad}-\nabla_{d}\delta g_{ba}))+\frac{1}{2}g_{fh}\nabla_{b}\nabla_{a}\delta g^{fh}\Bigg]+A_{a}\delta j^{a}\Bigg\}\nonumber\\
&+&\frac{1}{2}\int_{\mathcal{M}}d^{4}x\sqrt{-g}g^{ab}(\phi RA^{2}+\phi_{0}R_{cd}A^{c}A^{d})\delta g_{ab}.
\end{eqnarray}
The minimal coupling of the Maxwell field to the charge current densities of fluid $X$ gives a modification of the conjugate momentum, namely \cite{And5},
\begin{eqnarray}
\tilde{\mu_{a}}^{X}&=&\mu^{X}_{a}+e_{X}A_{a}.
\end{eqnarray}
The field equations obtained from the variation before above cannot be the final form, since the term proportional to $\delta n^{a}_{X}$ implies that the momentum $\tilde{\mu_{a}}^{X}$ must vanish \cite{And5}. This is essentially the condition that there be no energy present but clearly this is not viable \cite{And5}. This happens because the components of $\delta n^{a}_{X}$ cannot all be varied independently. A constrained variation is needed. A set of alternative variables which does precisely that is provided by the pull-back formalism-the $X^{A}_{X}$ can be varied independently, where $X^{A}_{X}$ ($A=(1, 2, 3)$) are coordinates of the $X$-fluid matter space. We also need to incorporate the fact that the fluid momenta have changed from $\mu^{X}_{a}$ to $\tilde{\mu_{a}}^{X}$ \cite{And5}. This means that we need to use the pull-back formalism for a single fluid approximation. For references and detailed discussion on the pull-back approach please check \cite{And5}, \cite{Tim13}, \cite{Tim14}, and \cite{Tim15}. The equations of motion for a general relativistic fluid are obtained from an action principle. This will form the foundation for the variations of the fundamental fluid variables in the action principle. After using the pull-back approach \cite{And5}, the equations of motion can be derived from the action principle. Therefore, a first-order variation of the fluid Lagrangian plus other variations of other Lagrangians that make up the total variation of equation (17) results in
\begin{eqnarray}
\delta S_{T}=~~~~~~~~~~~~~~~~~~~~~~~~~~~~~~~~~~~~~~~~~~~~~~~~~~~~~~~~~~~~~~~~~~~~~~~~~~~~~~~~~~~~~~~~~~~~~~~~~~~~~~~\nonumber\\\int_{\mathcal{M}}d^{4}x\sqrt{-g}\Bigg\{[\mu^{X}_{a}+e_{X}A_{a}]\delta n^{a}_{X}+\frac{1}{4\pi}[\nabla_{b}F^{ba}+4\pi(J^{a}_{X}+j^{a})+8\pi\phi RA^{a}+8\pi\phi_{0}R^{a}_{b}A^{b}]\delta A_{a}\nonumber\\+\frac{1}{2}\Bigg[(\psi\delta^{a}_{c}
+n^{a}_{X}\mu^{X}_{c})g^{cb}-\frac{1}{16\pi}(F_{cd}F^{cd}g^{ab}-4F^{ac}F^{b}_{c})-\phi g^{ah}g^{fb}R_{fh}A^{2}\Bigg]\delta g_{ab}\nonumber\\+2\phi g^{ab}A^{2}\Bigg[\frac{1}{2}\nabla_{c}(g^{cd}(\nabla_{a}\delta g_{db}+\nabla_{b}\delta g_{ad}
-\nabla_{d}\delta g_{ba}))+\frac{1}{2}g_{fh}\nabla_{b}\nabla_{a}\delta g^{fh}\Bigg]\nonumber\\+\frac{1}{2}g^{ab}[\phi RA^{2}+\phi_{0}R_{cd}A^{c}A^{d}]\delta g_{ab}-\mathcal{F}^{X}_{b}\xi^{b}_{X}+A_{a}\delta j^{a}\Bigg\},\nonumber\\
\end{eqnarray}
where $\mathcal{F}^{X}_{b}$ is the force density given by \cite{And5}
\begin{eqnarray}
\mathcal{F}^{X}_{b}&=&n^{a}_{X}\mathcal{W}^{X}_{ab}.
\end{eqnarray}
$\mathcal{W}^{X}_{ab}$ is defined as \cite{And5}
\begin{eqnarray}
\mathcal{W}^{X}_{ab}&=&2\nabla_{[a}\mu^{X}_{b]}\nonumber\\
&=&\nabla_{a}\mu^{X}_{b}-\nabla_{b}\mu^{X}_{a},
\end{eqnarray}
and $\psi$ is defined to be \cite{And5}
\begin{eqnarray}
\psi\equiv\Lambda-n^{a}_{X}\mu^{X}_{a}.
\end{eqnarray}

 For $S_{T}$ to be an extremum, the coefficients of $\delta A_{a}$ of equation (19) demand
\begin{eqnarray}
\nabla_{b}F^{ab}-8\pi\phi RA^{a}-8\pi\phi_{0} R^{a}_{b}A^{b}&=&4\pi J^{a}_{T}
\end{eqnarray}
including also equation (2), which are the equations of motion. For equation (2), symmetry of Christoffel symbols, $\Gamma^{d}_{ab}=\Gamma^{d}_{ba}$, enables us to substitute usual derivatives instead of covariant ones. Indeed,
\begin{eqnarray}
\nabla_{a}F_{bc}+\nabla_{b}F_{ca}+\nabla_{c}F_{ab}&=&0
\end{eqnarray}
means that
\begin{eqnarray}
\partial_{a}F_{bc}-\Gamma^{d}_{ab}F_{dc}-\Gamma^{d}_{ac}F_{bd}+\partial_{b}F_{ca}-\Gamma^{d}_{bc}F_{da}-\Gamma^{d}_{ba}F_{cd}+\partial_{c}F_{ab}-\Gamma^{d}_{ca}F_{db}-\Gamma^{d}_{cb}F_{ad}&=&0.
\end{eqnarray} 
Rewriting equation (25), we get,
\begin{eqnarray}
\partial_{a}F_{bc}-\Gamma^{d}_{ab}F_{dc}-\Gamma^{d}_{ac}F_{bd}+\partial_{b}F_{ca}-\Gamma^{d}_{bc}F_{da}+\Gamma^{d}_{ba}F_{dc}+\partial_{c}F_{ab}+\Gamma^{d}_{ca}F_{bd}+\Gamma^{d}_{cb}F_{da}&=&0.
\end{eqnarray}
This yields
\begin{eqnarray}
\partial_{a}F_{bc}+\partial_{b}F_{ca}+\partial_{c}F_{ab}&=&0
\end{eqnarray}
where all spatial derivatives are with respect to the comoving coordinates \cite{And1}. The formalism we used before above to derive the equations of motion for fluid $X$ is for a single-fluid approximation. In the next section, we will use a two-fluid formalism to derive the equations of motion for a two-fluid system but before doing so we will discuss briefly the action principle for a two-conducting fluid approximation.

\section{\label{sec4} The action principle for a two-conducting fluid approximation and modified Maxwell equations} 

Now, the fluid action $S_{M}$ has as its Lagrangian the master function $\Lambda$, which depends on the $n^{2}_{X}=-n^{X}_{a}n^{a}_{X}$, $n^{2}_{Y}=-n^{Y}_{b}n^{b}_{Y}$ and the $n^{2}_{XY}=-g_{ab}n^{a}_{X}n^{b}_{Y}$, where $n^{a}_{X}$ or $n^{b}_{Y}$ is the number density four current or flux whose magnitude $n_{X}$ or $n_{Y}$ respectively is the particle number density (see \cite{And5}), $X$ is a label for one fluid while $Y$ is a label for a different fluid from fluid $X$ and $g_{ab}$ is the metric. Ignoring boundary terms throughout, an arbitrary variation of $S_{M}$ with respect to the fluxes $n^{a}_{X}$, $n^{b}_{Y}$ and the metric results in \cite{And5}
\begin{eqnarray}
\delta S_{M}&=&\delta\int_{\mathcal{M}}d^{4}x\sqrt{-g}\Lambda
=\int_{\mathcal{M}}d^{4}x\sqrt{-g}\Bigg\{\sum_{i=\{X, Y\}}\mu^{i}_{a}\delta n^{a}_{i}+\frac{1}{2}[\Lambda g^{ab}+\sum_{i=\{X, Y\}}n^{a}_{i}\mu^{b}_{i}]\delta g_{ab}\Bigg\},\nonumber\\
\end{eqnarray}
where $\mathcal{M}$ is the manifold or hypersurface, $g$ is the determinant of the metric, and $\mu^{i}_{a}$ are the canonically conjugate momenta to $n^{a}_{i}$; that is, letting \cite{And5,Tim8}
\begin{eqnarray}
\mathcal{B}^{X}&=&-2\frac{\partial\Lambda}{\partial n^{2}_{X}}, \nonumber\\ \mathcal{B}^{Y}&=&-2\frac{\partial\Lambda}{\partial n^{2}_{Y}}, \nonumber\\ \mathcal{A}^{XY}=\mathcal{A}^{YX}&=&-\frac{\partial\Lambda}{\partial n^{2}_{XY}},
\end{eqnarray}
then \cite{Tim8}
\begin{eqnarray}
\mu^{X}_{a}&=&g_{ab}\{\mathcal{B}^{X}n^{b}_{X}+\sum_{Y\neq X}\mathcal{A}^{XY}n^{b}_{Y}\}, \nonumber\\
\mu^{Y}_{a}&=&g_{ab}\{\mathcal{B}^{Y}n^{b}_{Y}+\sum_{X\neq Y}\mathcal{A}^{YX}n^{b}_{X}\},
\end{eqnarray}
(see \cite{Leo}) where $XY$ incorporates entrainment between the fluids. 

 The Maxwell action is given by \cite{Leo} 
\begin{eqnarray}
S_{Max}&=&\frac{1}{16\pi}\int_{\mathcal{M}}d^{4}x\sqrt{-g}F_{ab}F^{ab},
\end{eqnarray}
and varying this action with respect to the vector potential $A_{a}$ and the metric $g_{ab}$ results in \cite{Leo}
\begin{eqnarray}
\delta S_{Max}&=&\frac{1}{4\pi}\int_{\mathcal{M}}d^{4}x\sqrt{-g}\{\nabla_{a}F^{ab}\}\delta A_{b}-\frac{1}{32\pi}\int_{\mathcal{M}}d^{4}x\sqrt{-g}(F_{cd}F^{cd}g^{ab}-4F^{ac}F^{b}_{c})\delta g_{ab}.
\end{eqnarray}
Since we are considering a two-conducting fluid (or multi-conducting fluid), the minimal coupling of the Maxwell field to the charge current densities is obtained from
\begin{eqnarray}
S_{C}&=&\int_{\mathcal{M}}d^{4}x\sqrt{-g}(\sum_{i=\{X, Y\}}J^{a}_{i}+j^{a})A_{a},
\end{eqnarray}
where $J^{a}_{i}$ are the flux currents. Varying action (33) with respect to $n^{a}_{i}$, $A_{a}$, $g_{ab}$ and $j^{a}$ results in
\begin{eqnarray}
\delta S_{C}&=&\int_{\mathcal{M}}d^{4}x\sqrt{-g}\sum_{i=\{X, Y\}}\Bigg\{[J^{a}_{i}+j^{a}]\delta A_{a}+e_{i}A_{a}\delta n^{a}_{i}+A_{a}\delta j^{a}+\frac{1}{2}[J^{a}_{i}+j^{a}]A_{a}g^{bc}\delta g_{bc}\Bigg\}.\nonumber\\
\end{eqnarray}
For a given system, the total action will thus become
\begin{eqnarray}
\delta S&=&\delta S_{M}+\delta S_{Max}+\delta S_{C}\nonumber\\
&=&\int_{\mathcal{M}}d^{x}\sqrt{-g}\Bigg\{\sum_{i=\{X, Y\}}[\mu^{i}_{a}+e_{i}A_{a}]\delta n^{a}_{i}+\frac{1}{4\pi}[\nabla_{b}F^{ba}+4\pi(\sum_{i=\{X, Y\}}J^{a}_{i}+j^{a})]\delta A_{a}\nonumber\\&+&\frac{1}{2}\Bigg[\Lambda g^{ab}
+\sum_{i=\{X, Y\}}[n^{a}_{i}\mu^{b}_{i}+(J^{c}_{i}+j^{c})A_{c}g^{ab}]-\frac{1}{16\pi}(F_{cd}F^{cd}g^{ab}-4F^{ac}F^{b}_{c})\Bigg]\delta g_{ab}+A_{a}\delta j^{a}\Bigg\}.\nonumber\\
\end{eqnarray}
Adding terms of the form $RA^{2}+R_{ab}A^{a}A^{b}$ results in the total action taking the form below,
\begin{eqnarray}
\delta S_{T}&=&\delta S_{M}+\delta S_{Max}+\delta S_{C}+\delta S_{\phi_{T}},
\end{eqnarray}
and writing explicitly we have
\begin{eqnarray}
\delta S_{T}&=&\nonumber\\&&\int_{\mathcal{M}}d^{4}x\sqrt{-g}\Bigg\{\sum_{i=\{X, Y\}}[\mu^{i}_{a}+e_{i}A_{a}]\delta n^{a}_{i}+\frac{1}{4\pi}[\nabla_{b}F^{ba}+4\pi\sum_{i=\{X, Y\}}(J^{a}_{i}+j^{a})\nonumber\\&+&8\pi\phi RA^{a}+8\pi\phi_{0} R^{a}_{b}A^{b}]\delta A_{a}
+\frac{1}{2}\Bigg[\Lambda g^{ab}+\sum_{i=\{X, Y\}}(n^{a}_{i}\mu^{b}_{i}+(J^{c}_{i}+j^{c})A_{c}g^{ab})\nonumber\\&-&\frac{1}{16\pi}(F_{cd}F^{cd}g^{ab}-4F^{ac}F^{b}_{c})-\phi g^{ha}g^{fb}R_{fh}A^{2}\Bigg]\delta g_{ab}\nonumber\\
&+&2\phi g^{ab}A^{2}\Bigg[\frac{1}{2}\nabla_{c}(g^{cd}(\nabla_{a}\delta g_{db}+\nabla_{b}\delta g_{ad}-\nabla_{d}\delta g_{ba}))+\frac{1}{2}g_{fh}\nabla_{b}\nabla_{a}\delta g^{fh}\Bigg]\nonumber\\
&+&\frac{1}{2}g^{ab}(\phi RA^{2}+\phi_{0}R_{cd}A^{c}A^{d})\delta g_{ab}+A_{a}\delta j^{a}\Bigg\}.
\end{eqnarray}

Similarly, as in the single-fluids approximation case, the minimal coupling of the Maxwell field to the charge current densities of the coupled fluids of $X$ and $Y$ gives a modification of the conjugate momentum, namely \cite{And5},
\begin{eqnarray}
\tilde{\mu}^{i}_{a}&=&\mu^{i}_{a}+e_{i}A_{a}.
\end{eqnarray}
Again and similarly the field equations obtained from the final variation before above cannot be the final form, since the term proportional to $\delta n^{a}_{i}$ implies that the momenta $\tilde{\mu}^{i}_{a}$ must vanish \cite{And5}. This is essentially the condition that there be no energy present but clearly this is not viable. This happens because the components of $\delta n^{i}_{X}$ cannot all be varied independently \cite{And5}. A constrained variation is needed. A set of alternative variables which does precisely that is provided by the pull-back formalism-the $X^{A}_{i}$ can be varied independently. We also need to incorporate the fact that the fluid momenta have changed from $\mu^{i}_{a}$ to $\tilde{\mu}^{i}_{a}$ \cite{And5}. This means that we need to use the pull-back formalism for two-fluids approximation. For a detailed discussion the reader is referred to \cite{Tim13, Tim14, Tim15, And5}. The equations of motion for a general relativistic fluid are obtained from an action principle. This will form the foundation for the variations of the fundamental fluid variables in the action principle. After using the pull-back approach for a two-fluid model \cite{And5}, the equations of motion can be derived from the action principle. Therefore, a first-order variation of the fluid Lagrangian plus other variations of the other Lagrangians that make up the total variation of equation (37) results in
\begin{eqnarray}
\delta S_{T}&=&\int_{\mathcal{M}}d^{4}x\sqrt{-g}\Bigg\{\sum_{i=\{X, Y\}}[(\mu^{i}_{a}+e_{i}A_{a})\delta n^{a}_{i}]+\frac{1}{4\pi}[\nabla_{b}F^{ba}+4\pi(\sum_{i=\{X, Y\}}J^{a}_{i}+j^{a})+8\pi\phi RA^{a}\nonumber\\
&+&8\pi\phi_{0}R^{a}_{b}A^{b}]\delta A_{a}+\frac{1}{2}\Bigg[(\psi\delta^{a}_{c}+\sum_{i=\{X, Y\}}n^{a}_{i}\mu^{i}_{c})g^{cb}-\frac{1}{16\pi}(F_{cd}F^{cd}g^{ab}-4F^{ac}F^{b}_{c})\nonumber\\
&-&\phi g^{ah}g^{fb}R_{fh}A^{2}\Bigg]\delta g_{ab}+2\phi g^{ab}A^{2}\Bigg[\frac{1}{2}\nabla_{c}(g^{cd}(\nabla_{a}\delta g_{db}+\nabla_{b}\delta g_{ad}-\nabla_{d}\delta g_{ba}))+\frac{1}{2}g_{fh}\nabla_{b}\nabla_{a}\delta g^{fh}\Bigg]\nonumber\\
&+&\frac{1}{2}[\phi RA^{2}+\phi_{0}R_{cd}A^{c}A^{d}]\delta g_{ab}-\sum_{i=\{X, Y\}}\mathcal{F}^{i}_{b}\xi^{b}_{i}\Bigg\}
\end{eqnarray}
where $\mathcal{F}^{i}_{b}$ is as defined in equation (20) except that the individual velocities are no longer parallel and $\psi$ is given by \cite{And5}
\begin{eqnarray}
\psi&=&\Lambda-\sum_{i=\{X, Y\}}n^{b}_{i}\mu^{i}_{b}.
\end{eqnarray}

Similarly, by varying $A_{a}$, modified Maxwell equations are obtained as in the single-fluids approximation case. For $S_{T}$ to be an extremum, the coefficients of $\delta A_{a}$ of equation (39) demand
\begin{eqnarray}
\nabla_{b}F^{ab}-8\pi\phi RA^{a}-8\pi\phi_{0} R^{a}_{b}A^{b}&=&4\pi(\sum_{i=\{X, Y\}}J^{a}_{i}+j^{a})
\end{eqnarray}
including also equation (2), which are the equations of motion. Similarly, as in the single-fluids approximation case, for equation (2), symmetry of Christoffel symbols, $\Gamma^{d}_{ab}=\Gamma^{d}_{ba}$, enables us to rewrite it in the form below,
\begin{eqnarray}
\partial_{a}F_{bc}+\partial_{b}F_{ca}+\partial_{c}F_{ab}&=&0
\end{eqnarray}
where all spatial derivatives are with respect to the comoving coordinates \cite{And1}.

Note that, even though the particular system we concentrate on consists of only two fluids, it illustrates all new features of a general multi-fluid system. We will scrutinise the effect of adding terms of the form $RA^{2}+R_{ab}A^{a}A^{b}$ first, to equations (12) and (35). The greatest step is to go from one to two fluids conceptually. Once this is done, a generalization to a system with more degrees of freedom is straightforward. Now, in order to study the evolution of magnetic fields we need to rewrite the Maxwell tensors in terms of the magnetic flux strength. We assume that magnetic evolution evolves through $RD$ , matter-radiation equality and $MD$ epochs as these epochs are much more prolonged than the inflation and reheating epochs. Therefore, the evolution of magnetic fields in the assumed epochs would change much more appreciably than in the epochs of inflation and reheating. We will do this in the sections to follow. We will start with the evolution of magnetic fields during the expansion of the Universe driven by the fluid of radiation. Therefore, we will use the equations of motion from the single-fluid formalism before above.

\section{Magnetic fields evolution during the $RD$ , matter-radiation equality and $MD$ epochs}

In this section we consider the evolution of magnetic fields during the $RD$ epoch first, then matter-radiation equality epoch and finally the $MD$ epoch. We proceed in the following subsections.

\subsection{$RD$ epoch}

The formalisms (both single-fluid and multi-fluid formalisms) we used to derive the Maxwell tensors can handle a number of different fluids other than just one or two \cite{Tim12}. In the cases we will be considering, electromagnetism is incorporated in the actions of the formalisms, thus allowing for plasmas and their effects on the systems. We consider equations (23) and (27). To study these equations we write them in terms of the $\vec{E}$ and $\vec{B}$ flux strength. The results will be vectors. The matrix we will use to enable us write equations (23) and (27) in terms of $\vec{E}$ and $\vec{B}$ flux strength is given below, 
\[\begin{bmatrix}F_{ab}\end{bmatrix}=\begin{bmatrix}
0 & -a^{2}E_{x} & -a^{2}E_{y} & -a^{2}E_{z} \\
a^{2}E_{x} & 0 & a^{2}B_{z} & -a^{2}B_{y} \\
a^{2}E_{y} & -a^{2}B_{z} & 0 & a^{2}B_{x} \\
a^{2}E_{z} & a^{2}B_{y} & -a^{2}B_{x} & 0 \\
\end{bmatrix}\]
where $[F_{ab}]=F_{ab}$ \cite{And1}. Considering equation (27), using $F_{ab}$ and the fact that $R=6\frac{\ddot{a}}{a^{3}}$ where $a$ is the cosmic scale factor, we recast equation (27) as \cite{And1}
\begin{eqnarray}
\frac{1}{a^{2}}\frac{\partial a^{2}\vec{B}}{\partial\eta}+\nabla\times\vec{E}&=&0
\end{eqnarray}
in vector form. Using $F_{ab}$, $R^{i}_{i}=\frac{\ddot{a}}{a^{3}}+\Bigg[\frac{\dot{a}}{a^{2}}\Bigg]^{2}$ (no sum on $i$) and $R=\frac{6\ddot{a}}{a^{3}}$ \cite{And1} we can recast equation (23) as
\begin{eqnarray}
\frac{1}{a^{2}}\frac{\partial(a^{2}\vec{E})}{\partial\eta}-\nabla\times\vec{B}-\frac{n}{\eta^{2}}\frac{\vec{A}}{a^{2}}&=&4\pi(\vec{J_{X}}+\vec{j})
\end{eqnarray}
in vector form where
\begin{eqnarray}
n&=&\eta^{2}\Bigg[8\pi\Bigg(6\phi\frac{\ddot{a}}{a}+\phi_{0}\Bigg(\frac{\ddot{a}}{a}+\Bigg(\frac{\dot{a}}{a}\Bigg)^{2}\Bigg)\Bigg)\Bigg].
\end{eqnarray}
$\phi$ and $\phi_{0}$ are coupling constants and $n$ is a constant whenever $a(\eta)$ varies as a power of $\eta$. In equation (44) we have $\vec{J}_{X}$ which is the current due to fluid $X$ representing the $RD$ epoch (as a fluid) and $\vec{j}$ is the current due to plasma. Therefore, we have
\begin{eqnarray}
\vec{J}_{X}+\vec{j}&=&\vec{J}_{T}
\end{eqnarray}
where $\vec{J}_{T}$ is the total current. Due to dominant contribution of plasma effects to flux current, $\vec{J}_{X}$ or mathematically, when 
$|\vec{J_{T}}|=J_{T}=\sqrt{J^{2}_{X}+J^{2}-2J_{X}J\cos\theta}=J\sqrt{1+\frac{J^{2}_{X}}{J^{2}}-\frac{2J_{X}\cos\theta}{J}}\approx J\Bigg(1+\frac{J^{2}_{X}}{2J^{2}}-\frac{J_{X}\cos\theta}{J})+\dots\Bigg)$ 
to linear order (as the higher-order terms are very minute or negligible) implying that if
\begin{eqnarray}
|\vec{j}|^{2}\gg|\vec{J}_{X}|^{2}
\end{eqnarray}
(note that $J\gg J_{X}\cos\theta$) we have $J_{T}\approx J$ hence, $|\vec{J_{T}}|\approx|\vec{J}|$ and then
\begin{eqnarray}
\vec{J}_{T}\approx\vec{j}.
\end{eqnarray}
For further clarification please refer to appendix $A$. This then implies that equation (44) can be rewritten as
\begin{eqnarray}
\frac{1}{a^{2}}\frac{\partial(a^{2}\vec{E})}{\partial\eta}-\nabla\times\vec{B}-\frac{n}{\eta^{2}}\frac{\vec{A}}{a^{2}}&=&4\pi\vec{j}
\end{eqnarray}
in vector form. We can now study the evolution of magnetic fields in the $RD$ epoch of the Universe using the generalised cosmological Ohm's law where $\vec{j}=\vec{J}$. But before doing so we will rescale the electromagnetic fields, currents and the electromagnetic four vector potential. This is possible due to conformal invariance \cite{Tim26}. The equations that result from this are \cite{Tim26}
\begin{eqnarray}
\tilde{E}\equiv a^{2}\vec{E} , \nonumber\ \tilde{B}\equiv a^{2}\vec{B} , \nonumber\ \tilde{J}\equiv a^{3}\vec{J} , \nonumber\
\tilde{J}_{r}\equiv a^{3}\vec{J}_{r} , \nonumber\ \tilde{A}\equiv a\vec{A}.
\end{eqnarray}
We now rewrite equation (49) in the form below,
\begin{eqnarray}
\frac{\partial(a^{2}\vec{E})}{\partial\eta}-\nabla\times a^{2}\vec{B}-\frac{n}{\eta^{2}a^{2}}a^{2}\vec{A}&=&\frac{4\pi a^{3}\vec{J}}{a}
\end{eqnarray}
as a vector equation. Simplifying yields,
\begin{eqnarray}
\frac{\partial\tilde{E}}{\partial\eta}-\nabla\times\tilde{B}-\frac{na\tilde{A}}{\eta^{2}a^{2}}&=&\frac{4\pi\tilde{J}}{a}.
\end{eqnarray}
We then take the curl of equation (51) and this yields,
\begin{eqnarray}
\frac{\partial\nabla\times\tilde{E}}{\partial\eta}-\nabla\times\nabla\times\tilde{B}-\frac{n}{\eta^{2}a^{2}}\tilde{\nabla}\times\tilde{A}&=&\frac{4\pi}{a}\nabla\times\tilde{J}
\end{eqnarray}
where $\tilde{\nabla}=a\nabla$. But equation (43) can be written in the form below \cite{Tim26},
\begin{eqnarray}
\frac{\partial a^{2}\vec{B}}{\partial\eta}+\nabla\times a^{2}\vec{E}&=&0,
\end{eqnarray}
and simplifying \cite{Tim26},
\begin{eqnarray}
\frac{\partial\tilde{B}}{\partial\eta}+\nabla\times\tilde{E}&=&0.
\end{eqnarray}
Using equation (54), the identities
\begin{eqnarray}
\nabla\times\nabla\times\tilde{B}&=&\nabla(\nabla.\tilde{B})-\nabla^{2}\tilde{B} \nonumber\\
&=&a^{2}\nabla(\nabla.\vec{B})-\nabla^{2}\tilde{B} \nonumber\\
&=&0-\nabla^{2}\tilde{B} \nonumber\\
&=&-\nabla^{2}\tilde{B}
\end{eqnarray}
(since $\nabla a^{2}=0$) and
\begin{eqnarray}
\tilde{\nabla}\times\tilde{A}&=&\tilde{B}
\end{eqnarray}
in equation (52) results in
\begin{eqnarray}
\frac{\partial}{\partial\eta}\Bigg(-\frac{\partial\tilde{B}}{\partial\eta}\Bigg)+\nabla^{2}\tilde{B}-\frac{n}{\eta^{2}a^{2}}\tilde{B}&=&\frac{4\pi}{a}\nabla\times\tilde{J}
\end{eqnarray}
which can be rewritten in the form below,
\begin{eqnarray}
\frac{\partial^{2}\tilde{B}}{\partial\eta^{2}}-\nabla^{2}\tilde{B}+\frac{n}{\eta^{2}a^{2}}\tilde{B}&=&-\frac{4\pi}{a}\nabla\times\tilde{J}.
\end{eqnarray}
Now, equation (5) in rescaled form is \cite{Tim26}
\begin{eqnarray}
\tilde{J}&=&\hat{\sigma}\tilde{E}-\frac{\eta_{T}\hat{\sigma}\tilde{J}_{r}}{a}
\end{eqnarray}
where \cite{Tim26}
\begin{eqnarray}
\hat{\sigma}&=&a\sigma.
\end{eqnarray}

 Then from equation (59) we have \cite{Tim26}
\begin{eqnarray}
\nabla\times\tilde{J}&=&\hat{\sigma}\nabla\times\tilde{E}-\frac{\eta_{T}\hat{\sigma}}{a}\nabla\times\tilde{J}_{r}.
\end{eqnarray}
With equation (54), equation (61) can be rewritten in the form below \cite{Tim26},
\begin{eqnarray}
\nabla\times\tilde{J}&=&-\hat{\sigma}\frac{\partial\tilde{B}}{\partial\eta}-\frac{\eta_{T}\hat{\sigma}}{a}\nabla\times\tilde{J}_{r}.
\end{eqnarray}
Using equation (62), equation (58) can be rewritten in the form
\begin{eqnarray}
\frac{\partial^{2}\tilde{B}}{\partial\eta^{2}}-\nabla^{2}\tilde{B}+\frac{n}{\eta^{2}a^{2}}\tilde{B}&=&-\frac{4\pi}{a}\Bigg(-\hat{\sigma}\frac{\partial\tilde{B}}{\partial\eta}-\frac{\eta_{T}\hat{\sigma}}{a}\nabla\times\tilde{J}_{r}\Bigg).
\end{eqnarray}
Rewriting equation (63) in a more suitable form results in
\begin{eqnarray}
\tilde{B^{\prime\prime}}-\nabla^{2}\tilde{B}+\frac{n}{\eta^{2}a^{2}}\tilde{B}-\frac{4\pi\hat{\sigma}}{a}\tilde{B^{\prime}}&=&\frac{4\pi\eta_{T}\hat{\sigma}}{a^{2}}\nabla\times\tilde{J}_{r}
\end{eqnarray}
where primes denote derivatives with respect to conformal time. In \cite{And1} $\frac{1}{\eta a}\sim H$. But $H\sim R^{\frac{1}{2}}$, $m_{\gamma}\sim R^{\frac{1}{2}}$, \cite{And1} where $m_{\gamma}$ is the photon mass. This implies that $\frac{1}{\eta a}\sim m_{\gamma}$ \cite{And1}. Therefore, equation (64) will become
\begin{eqnarray}
\tilde{B^{\prime\prime}}-\nabla^{2}\tilde{B}+m^{2}_{\gamma}n\tilde{B}-\frac{4\pi\hat{\sigma}}{a}\tilde{B^{\prime}}&=&\frac{4\pi\eta_{T}\hat{\sigma}}{a}\nabla\times\tilde{J}_{r}.
\end{eqnarray}
On comoving scales larger than $\hat{\sigma}^{-1}$ one can conclude that the first three ($3$) terms on the left hand side ($LHS$) can be neglected with respect to the fourth term (check condition of coupling constants $\phi_{0}$ and $\phi$ just after equation (74) showing that neglect is reasonable before above). This is equivalent to dropping $\vec{J}$ with respect to $\sigma_{E}\vec{E}$ in equation (5). This implies that equation (61) becomes \cite{Tim26}
\begin{eqnarray}
\frac{\eta_{T}\hat{\sigma}}{a}\nabla\times\tilde{J}_{r}&=&\hat{\sigma}\nabla\times\tilde{E}.
\end{eqnarray}
Then using equation (66) in equation (65), we have
\begin{eqnarray}
\tilde{B^{\prime\prime}}-\nabla^{2}\tilde{B}+m^{2}_{\gamma}n\tilde{B}-\frac{4\pi\hat{\sigma}}{a}\tilde{B^{\prime}}&=&\frac{4\pi\hat{\sigma}}{a}\nabla\times\tilde{E}.
\end{eqnarray}
Using equation (54) in equation (67) yields,
\begin{eqnarray}
\tilde{B^{\prime\prime}}-\nabla^{2}\tilde{B}+m^{2}_{\gamma}n\tilde{B}-\frac{4\pi\hat{\sigma}}{a}\tilde{B^{\prime}}&=&-\frac{4\pi\hat{\sigma}}{a}\tilde{B^{\prime}}.
\end{eqnarray}
This simplifies to
\begin{eqnarray}
\tilde{B^{\prime\prime}}-\nabla^{2}\tilde{B}+m^{2}_{\gamma}n\tilde{B}&=&0,
\end{eqnarray}
at the start of the $RD$ epoch (or matter-radiation equality epoch or $MD$ epoch as we will find out later on in this paper for these epochs in brackets). This equation is similar to equation (4) in \cite{Tim32}. Equation (4) in \cite{Tim32} is a wavelike equation obeyed by the magnetic component of the Maxwell field or inflation-produced magnetic fields at the moment they exit the Hubble horizon in standard electromagnetism and in the absence of currents. Causality guarantees that there are no coherent electric currents with superhorizon correlations, which in turn implies that there is no magnetic flux-freezing on super-Hubble scales \cite{Tim200}. This means that the adiabatic decay-law is not guaranteed on superhorizon lengths \cite{Tim32,Tim201,Tim400,Tim202} as magnetic freeze-in cannot take place without the electric currents \cite{Tim203}. However, in this study, currents with both cosmological and superhorizon correlations exist meaning they are valid on both cosmological horizon and superhorizon scales as argued in \cite{Tim26}. The ideal-$MHD$ limit cannot be applied to super-Hubble scales as the process of magnetic-flux freezing itself is causal and causal physics can never affect superhorizon sized perturbations (it is the currents that eliminate the electric fields and freeze their magnetic counterparts into the matter and this is only possible within the Hubble horizon \cite{Tim32}).

We consider the simple Ohm's law for a charged or conducting fluid given as $\vec{J}_{q0}=\sigma_{c}\vec{E}$ to clarify the arguments before above where $\vec{J}_{q0}$ is the 3-current, $\sigma_{c}$ is the electrical conductivity of the medium (plasma) \cite{Tim204,Tim205} and $\vec{E}$ is the electric field. It is easy then to see that 3-current is related via Ohm's law before above. $\sigma_{c}\rightarrow\infty$, at the ideal magnetohydrodynamic ($MHD$) limit when the conductivity is very high. This means that the electric field vanishes and the currents keep the magnetic field frozen-in with the fluid \cite{Tim206}. On the other hand, when $\sigma_{c}\rightarrow0$ the currents vanish despite the presence of a finite electric field \cite{Tim206}. In this study we assume a simple Ohm's law for subhorizon scales throughout the post-inflationary Universe as in \cite{Tim32}.

We now consider the generalised cosmological Ohm's law given by $\vec{J}+\sigma_{E}\eta_{T}\vec{J}_{r}\simeq\sigma_{E}\vec{E}$ or equation (5). Similarly, at the ideal $MHD$ limit the conductivity is very high implying $\sigma_{E}\rightarrow\infty$. This means that the electric field should vanish and then the currents will keep the magnetic field frozen-in with the fluid. But after very careful scrutiny of the equation before above representing the generalised cosmological Ohm's law or equation (5), we see that when $\sigma_{E}\rightarrow\infty$ the electric field does not vanish. This implies that currents cannot keep the magnetic field frozen-in with the fluid. Hence, there should not be adiabatic decay from the beginning of the $RD$ epoch to much later during the $MD$ epoch or from the beginning of matter-radiation equality transit epoch to much later during the $MD$ epoch or from the beginning of the $MD$ epoch to much later during the $MD$ epoch (as we will find out later on in this paper for the latter two scenarios) on horizon (and superhorizon) scales. This refutes the conclusion in \cite{Tim26} that due to the sizeable conductivity of the plasma, modifications of the standard (adiabatic) evolution of magnetic fields are severely limited on cosmological horizon (and superhorizon) scales which implies that any departure from magnetic flux freezing behaviour is inhibited. Again, whether there are currents or no currents, magnetic flux-freezing cannot take place on horizon (and super-Hubble) scales as argued before above (this is true after using the generalised cosmological Ohm's law where the effective current is the photon current, $\vec{J}_{r}$ and this will be clarified later in this paper). On the other hand, when $\sigma_{E}\rightarrow0$, not all currents vanish and these currents are the external sources of the surviving electric field. At $z\sim300$ the photon current ceases to be an active source \cite{Tim26} (assuming that there are currents on superhorizon scales before $z\sim300$ although in this study, the generalised cosmological Ohm's law is valid on horizon scales only and not superhorizon scales implying that there are no currents on superhorizon scales as in \cite{Tim32}). The effective photon current becomes irrelevant for $z\lesssim300$ \cite{Tim26}. The evolution of inflation-produced magnetic fields will continue on superhorizon scales until much later during the $MD$ epoch when they cross the horizon (for a second time) and go back to adiabatic magnetic decay until the present time (this is after assuming a simple Ohm's law for $z\lesssim300$ for subhorizon scales and assuming that there were currents before $z\sim300$ on superhorizon scales). However, in this study again, there are no currents on superhorizon scales as in \cite{Tim32} even though their presence will not result in adiabatic decay of inflation-produced magnetic fields as argued before above but there are currents on the horizon scales only, represented by the generalised cosmological Ohm's law where the effective current is the photon current and their presence are justified by the reasons before above and is clarified by the equations and arguments before above and later in this paper.

Therefore, in this study, the above equation or equation (69) (and/ or for the case when $n=0$ below or later in this paper) above is obeyed by the magnetic component of the modified Maxwell field or inflation-produced magnetic fields as they cross the Hubble horizon at the beginning of the $RD$ epoch (or matter-radiation equality epoch or $MD$ epoch as we will find out later on for these epochs mentioned in brackets), and in the presence of currents. This is logical given the arguments before above. Now to solve the above equation or equation (69) we introduce the harmonic splitting of $\tilde{B}=\Sigma_{k}\tilde{B}_{k}Q^{k}$ where $\tilde{B}_{k}$ is the $k^{th}$ magnetic mode and $k$ is the associated comoving eigenvalue \cite{Tim400}. $Q^{k}$ are pure-vector harmonics that satisfy the conditions $Q^{\prime k}=0=\nabla Q^{k}$ and the other version of the Laplace-Beltrami equation, that is \cite{Tim400}
\begin{eqnarray}
\nabla^{2}Q^{k}&=&-k^{2}Q^{k}.
\end{eqnarray}
Applying the above decomposition to equation (69), the harmonics decouple \cite{Tim400}, and the wave formula of the $k^{th}$ magnetic mode assumes the form
\begin{eqnarray}
\tilde{B^{\prime\prime}_{k}}+k^{2}\tilde{B_{k}}+m^{2}_{\gamma}n\tilde{B_{k}}&=&0.
\end{eqnarray}

We consider the case where $n\neq0$. Rewriting equation (71) in a more suitable form we have
\begin{eqnarray}
\tilde{B^{\prime\prime}_{k}}+\mathcal{q}^{2}\tilde{B_{k}}&=&0
\end{eqnarray}
where $\mathcal{q}^{2}=k^{2}+m^{2}_{\gamma}n$. Solving the above equation yields (note that $|\vec{B}_{k}|=B_{k}$)
\begin{eqnarray}
a^{2}B_{k}&=&C_{1}\cos\mathcal{q}\eta + C_{2}\sin\mathcal{q}\eta
\end{eqnarray}
for $\mathcal{q}=\sqrt{k^{2}+m^{2}_{\gamma}n}>0$ where $C_{1}$ and $C_{2}$ are constants and the equation is written for the actual magnetic field ($B=\frac{\tilde{B}}{a^{2}}$). This is possible for $k>0, \phi_{0}>0$ and $\phi>0$ which implies that $n>0$ ($n$, equation (45) is fixed and is a constant. $\phi$ and $\phi_{0}$, coupling constants of curvature and the electromagnetic field potential are in equation (45). $k$ is the associated comoving eigenvalue of the $k^{th}$ magnetic mode defined in the paragraph before equation (70)). This equation is similar to the equation in \cite{Tim32} that applies to or obeyed by the magnetic component of the Maxwell field or inflationary magnetic fields as they cross the horizon during the de Sitter era and in the absence of currents. But we are considering the $RD$ epoch or are in the $RD$ epoch. This equation then applies to or is obeyed by the magnetic component of the modified Maxwell field or inflation-produced magnetic fields as they exit the Hubble horizon at the beginning of the $RD$ epoch (or matter-radiation equality epoch or $MD$ epoch as we will find out later on when analyzing these epochs in brackets) and in the presence of currents. This equation is a differential equation that accepts an oscillatory solution where $\mathcal{q}\eta=\frac{\lambda_{H}}{\lambda_{\mathcal{q}}}$. Relative to the Hubble horizon ($\lambda_{H}=\frac{1}{H}$) where $H$ is the Hubble radius, the ratio ($\frac{\lambda_{H}}{\lambda_{\mathcal{q}}}$) measures the physical size of the magnetic mode ($\lambda_{\mathcal{q}}=\frac{a}{\mathcal{q}}$). So when the inflation-produced magnetic fields at the horizon initially are well outside the Hubble radius, that is for $\frac{\lambda_{H}}{\lambda_{\mathcal{q}}}\ll1$, meaning $\mathcal{q}\eta\ll1$ in conformal-time terms, a simple Taylor expansion reduces the above equation (73) to the power law
\begin{eqnarray}
a^{2}B_{k}&=&C_{1}+C_{2}\mathcal{q}\eta
\end{eqnarray}
where $a=a(\eta)$ and $\mathcal{q}=\sqrt{k^{2}+m^{2}_{\gamma}n}$ again. $k$, $\phi_{0}$ and $\phi$ are constants whose values are much less than 1 (that is $k\ll1$, $\phi_{0}\ll1$ and $\phi\ll1$ hence, $n\ll1$) and therefore, the condition $\mathcal{q}\eta\ll1$ is achieved ($n$ and $k$ are in $\mathcal{q}$ while the coupling constants $\phi$ and $\phi_{0}$ are in $n$). Note that the subscript $0$ on $\phi_{0}$ doesn't indicate a given initial time or today or present time. It is just zero (0) in order to differentiate $\phi_{0}$ from $\phi$.

We now consider the case where $n=0$. This case is for standard electromagnetism in \cite{Tim32} as we will find out now. Equation (69) reduces to $\tilde{B^{\prime\prime}}-\nabla^{2}\tilde{B}=0$, which is exactly equation (4) in \cite{Tim32}, a wavelike equation obeyed by the magnetic component of the Maxwell field or inflation-produced magnetic fields at the moment they exit the Hubble horizon in standard electromagnetism during the de Sitter era and in the absence of currents. Equation (71) will reduce to
\begin{eqnarray}
\tilde{B^{\prime\prime}_{k}}+k^{2}\tilde{B_{k}}&=&0
\end{eqnarray}
\cite{Tim100}. Solving the above equation yields
\begin{eqnarray}
a^{2}\vec{B}_{k}&=&C_{3}\cos k\eta+C_{4}\sin k\eta
\end{eqnarray}
where $C_{3}$ and $C_{4}$ are constants. Note that this equation can be obtained from equation (73) before above when $n=0$ too.

This equation is exactly the same as the equation that applies or is obeyed by the magnetic component of the Maxwell field or inflationary magnetic fields as they cross the horizon during the de Sitter era and in the absence of currents \cite{Tim32}. But we are considering the $RD$ epoch or are in the $RD$ epoch. This equation then applies to the magnetic component of the standard Maxwell field or inflation-produced magnetic fields as they cross the Hubble horizon at the beginning of the $RD$ epoch (or matter-radiation equality epoch or $MD$ epoch as we will find out later on when analyzing these epochs in brackets) and in the presence of currents. This equation is a differential equation that accepts an oscillatory solution where $k\eta=\frac{\lambda_{H}}{\lambda_{k}}$ \cite{Tim32}. Relative to the Hubble horizon ($\lambda_{H}=\frac{1}{H}$) where $H$ is the Hubble radius, the ratio ($\frac{\lambda_{H}}{\lambda_{k}}$) measures the physical size of the magnetic mode ($\lambda=\frac{a}{k}$) \cite{Tim32}. So when the inflation-produced magnetic fields at the horizon initially are well outside the Hubble radius, that is for $\frac{\lambda_{H}}{\lambda_{k}}\ll1$, meaning $k\eta\ll1$ in conformal-time terms, a simple Taylor expansion reduces the above equation (76) to the power law
\begin{eqnarray}
 a^{2}B_{k}&=&C_{3}+C_{4}k\eta
\end{eqnarray}
also found in \cite{Tim32} with $a=a(\eta)$ (please check the notes \cite{Tim403}).

Causality implies that the time required for the freezing-in information for the magnetic field to travel the whole length of a super-Hubble magnetic field is longer than the age of the Universe at the time \cite{Tim32}. Hence, the magnetic field cannot re-adjust itself to the new environment and freeze-in, until it has crossed back inside the horizon much later during the $MD$ epoch and come into full causal contact \cite{Tim32}. The magnetic field is immune to causal physics and retains only the memory of its distant past as long as it remains outside the Hubble horizon \cite{Tim32}. This means that the magnetic evolution is still governed by equations (74) and (77) in the presence of currents valid on the cosmological horizon (and superhorizon scales).
 
When we add terms of the form(s) $RA^{2}$ and/ or $R_{ab}A^{a}A^{b}$ in actions (12) and (35), we obtain equations similar to equations (74) and (77). The only difference is that for the third term in equation (64) we have for the constant $n$,
\begin{eqnarray}
n&=&n_{0}=\eta^{2}8\pi 6\phi\frac{\ddot{a}}{a}
\end{eqnarray}
where the case $RA^{2}$ is considered and
\begin{eqnarray}
n&=&n_{1}=\eta^{2}8\pi\phi_{0}\Bigg[\frac{\ddot{a}}{a}+\Bigg(\frac{\dot{a}}{a}\Bigg)\Bigg]
\end{eqnarray}
where the case $R_{ab}A^{a}A^{b}$ is considered and again we obtain equations similar to equations (74) and (77). We arrive at the same conclusions as for the case where we considered $RA^{2}+R_{ab}A^{a}A^{b}$. Inflation-produced, large-scale magnetic fields cross the Hubble horizon for the first time at the beginning of the $RD$, evolve superadiabatically until their second horizon crossing much later during the $MD$ epoch and then they go back to adiabatic decay until the present time. This implies that inflation-produced, large-scale magnetic fields evolved adiabatically before the beginning of the epoch of $RD$ and after they crossed the horizon for a second time. We now consider the evolution of magnetic fields in the matter-radiation equality epoch.

\subsection{Matter-radiation equality transit (epoch)}

As the Universe expands, its evolution is traditionally examined by monitoring how its material content evolves \cite{Tim300}. As an isolated system, this model is expressed as the equation of motion of the bulk but segmented into different epochs. Particularly, the evolution of the flat $FLRW$ Universe is separated into different epochs that are characterised by the dynamics of whichever mass-energy constituent is dominant at the time \cite{Tim300}. The standard analysis of the evolution of the Universe in a particular epoch usually considers the evolution of the dominant energy density only; not considering all others. Whereas this represents the limiting case, in principle the contributions from others cannot always be ignored. In particular, the proximity of the equality of the mass-energy density of the matter-radiation equality transit cannot be neglected. Therefore, the evolution of the total energy density rather than individual energy densities during the different epochs are examined in \cite{Tim300}. It is found that $\frac{1}{12}-\delta\leq w_{eff}\leq\frac{1}{6}-\delta$, where $\delta=\frac{\rho_{0}}{\rho_{Tot}}$. $\rho_{0}$ is a constituent of the total mass-energy density representing dynamical dark energy or other forms of mass-energy density, $\rho_{Tot}$ is the total density of the Universe at a particular time in its cosmic history and $w_{eff}$ is the effective equation of state ($EoS$) (and it can be used as the $EoS$ for particular matter types) during the epoch of matter-radiation equality transit. In general $\delta$ is comparatively small during the transition of matter-radiation equality and in standard analysis can be taken to be negligible. Nonetheless, as this is a simplifying assumption, one must take care to avoid mistakes. The effective $EoS$ for the total energy density has the potential of altering how fields evolve during the matter-radiation equality transit epoch. One example is the evolution of inflation-produced large-scale magnetic fields which may couple electrically to radiation and gravitationally to matter \cite{Tim300}. Neglecting how one component evolves would lead to an over-or under-estimation of the field strength. The matter-radiation equality transit epoch is much more prolonged than the inflation or reheating epochs hence, it would be logical to infer that $w$ is constant throughout this so-called transit \cite{Tim32}.

It is shown in \cite{Tim12} that it is possible to have a cosmological epoch where there is a relative flow of radiation with respect to the matter, but out of which the expansion becomes isotropic and the relative flow dissipates. This transit of matter-radiation equality (epoch) puts forward a Bianchi $I$ behaviour with a space-like privileged direction. This two-fluid nature of the problem introduces several terms that are not present in the one-fluid case. We introduce the so-called cross-constituent coupling, which occurs when the $EoS$ has terms containing both fluid densities. It is an equilibrium property and hence, non-dissipative. 
 
We then assume a general relativistic two-fluid model for a coupled system of matter (non-zero rest mass) and radiation (zero rest mass) \cite{Tim12}. The two fluids are allowed to interpenetrate and exhibit a relative flow with respect to each other (, implying, in general, an anisotropic Universe). The initial conditions used are such that the massless fluid flux dominates early on so that the situation is effectively that of a single-fluid and one has the usual $FLRW$ spacetime. This two-fluid model introduced is valid for applications in cosmology \cite{Tim12}.

The model developed in \cite{Tim12} is the simplest set-up of what might be envisioned for the transition itself, for we have not taken into account, for example, the non-conservation of the photon number through its coupling with luminous matter, matter flows with more than one constituent or relative flows at arbitrary angles \cite{Tim12}. One would not be too surprised if comparisons to observational data indicated the need for a more elaborate model. Here also, we have assumed the matter fluid to have only one flux-component. This assumption might not be reasonable; however, it is largely a scale-dependent statement.
 
The formalism developed can in principle handle a number of different fluids other than just two \cite{Tim12}. In the case we are considering, electromagnetism is incorporated in the action, thus allowing for plasmas and their effects on the system. Now the two-fluid cosmology we will consider has a combination of matter with mass $m^{Y}=m$, and radiation, which means that $m^{X}=0$. As mentioned before above we assume that there is cross-constituent coupling and zero entrainment \cite{Tim12}. We have particle flux of matter given by $n_{Y}=n$ and ignoring dissipation we have for the entropy flux of the system $n_{X}=s$, and the bulk of this is due to radiation. We set $\mu^{X}=T$, which is the temperature.

We now investigate how a cross-constituent term can come about. We consider the usual way of combining a (non-relativistic) gas and radiation in the energy density and pressure:
\begin{eqnarray}
\rho&=&mn+\frac{3}{2}nT+\alpha T^{4},
\end{eqnarray}
\begin{eqnarray}
p&=&nT+\frac{1}{3}\alpha T^{4},
\end{eqnarray}
where $\alpha$ is constant \cite{Tim12}. Taking as our fundamental thermodynamic variables $n$ and $s$ will result in $T=\frac{\partial\rho}{\partial s}$ which is a function of both. Therefore, the ideal gas contribution will generate a cross-constituent coupling and a measure of the interactions are the cross-constituent couplings defined as
\begin{eqnarray}
C_{YX}\equiv\frac{\partial\ln\mu^{Y}}{\partial\ln n_{X}}&=&\frac{\mu^{X}n_{X}}{\mu^{Y}n_{Y}}C_{XY}.
\end{eqnarray}
The $C_{YX}$ represents a key channel through which the two fluids see each other especially when the entrainment is zero \cite{Tim12}. 

We now consider equation (41) with the aim of finding out how magnetic fields evolve during the transit of matter-radiation equality (epoch). Rewriting it in more suitable form (tensor form still) we have
\begin{eqnarray}
\nabla_{b}F^{ab}-8\pi\phi RA^{a}-8\pi\phi_{0}R^{a}_{b}A^{b}&=&4\pi(J^{a}_{X}+J^{a}_{Y}+j^{a}) .
\end{eqnarray}
Now using $F_{ab}$, the fact that $R^{i}_{i}=\frac{\ddot{a}}{a^{3}}+\Bigg[\frac{\dot{a}}{a^{2}}\Bigg]^{2}$ (no sum on $i$) and $R=6\frac{\ddot{a}}{a^{3}}$ \cite{And1}, we can recast equation (83) as
\begin{eqnarray}
\frac{1}{a^{2}}\frac{\partial(a^{2}\vec{E})}{\partial\eta}-\nabla\times\vec{B}-\frac{n}{\eta^{2}}\frac{\vec{A}}{a^{2}}&=&4\pi(\vec{J}_{X}+\vec{J}_{Y}+\vec{j})
\end{eqnarray}
in vector form where $\frac{n}{\eta^{2}}$ is equation (45).
\begin{eqnarray}
\vec{J_{Z}}+\vec{j}&=&\vec{J}_{T}
\end{eqnarray}
where $\vec{J}_{T}$ is the total current and $\vec{J_{Z}}=\vec{J_{X}}+\vec{J_{Y}}$. Due to dominant contribution of plasma effects to that of the total of flux currents, $\vec{J}_{X}+\vec{J}_{Y}$ or mathematically, when $|\vec{J_{T}}|=J_{T}=\sqrt{J^{2}_{Z}+J^{2}-2J_{Z}J\cos\theta}=J\sqrt{1+\frac{J^{2}_{Z}}{J^{2}}-\frac{2J_{Z}\cos\theta}{J}}\approx J\Bigg(1+\frac{J^{2}_{Z}}{2J^{2}}-\frac{J_{Z}\cos\theta}{J}+\dots\Bigg)$ to linear order (as the higher-order terms are very minute or negligible), where $\frac{J^{2}_{Z}}{J^{2}}=\frac{J^{2}_{X}+J^{2}_{Y}}{J^{2}}-\frac{2J_{X}J_{Y}\cos\theta}{J^{2}}$ or $\frac{J_{Z}}{J}=\sqrt{\frac{J^{2}_{X}+J^{2}_{Y}}{J^{2}}-\frac{2J_{X}J_{Y}\cos\theta}{J^{2}}}$, and then if
\begin{eqnarray}
|\vec{j}|^{2}\gg|\vec{J_{Z}}|^{2}
\end{eqnarray}
we have $J_{T}\approx J$ hence, $|\vec{J_{T}}|\approx|\vec{J}|$ implying that
\begin{eqnarray}
\vec{J_{T}}\approx\vec{j}.
\end{eqnarray}
For further clarification please refer to appendix $A$. We can then rewrite equation (84) as
\begin{eqnarray}
\frac{1}{a^{2}}\frac{\partial(a^{2}\vec{E})}{\partial\eta}-\nabla\times\vec{B}-\frac{n}{\eta^{2}}\frac{\vec{A}}{a^{2}}&=&4\pi\vec{j}
\end{eqnarray}
in vector form.

We can see that equation (88) above is the same as equation (49). Therefore, from this point onwards [from equation (49)], we can use the same analysis as we did during the $RD$ epoch subsection. We will obtain the same equations and hence, we will have the same conclusions as in the analysis of the $RD$ epoch except that for this epoch, inflation-produced, large-scale magnetic fields will cross the Hubble horizon for the first time at the beginning of this epoch of matter-radiation equality transit, evolve superadiabatically until they cross the horizon for a second time much later during the epoch of $MD$ and then go back to adiabatic decay until the present time. This implies that before the beginning of this epoch of matter-radiation equality transit, inflation-produced, large-scale magnetic fields evolved adiabatically starting at the inflationary epoch and after they crossed the horizon for a second time much later during the $MD$ epoch. We now consider the evolution of inflation-produced, large-scale magnetic fields during the $MD$ epoch in the next subsection.
 
\subsection{$MD$ epoch}

The modelling or analysis of magnetic fields evolution during the $MD$ epoch is exactly the same as that during the $RD$ epoch up to equation (43). 

Therefore, then, using $F_{ab}$, $R^{i}_{i}=\frac{\ddot{a}}{a^{3}}+\Bigg(\frac{\dot{a}}{a}\Bigg)^{2}$ (no sum on $i$) and $R=6\frac{\ddot{a}}{a^{3}}$ \cite{And1} we can recast equation (23) as
\begin{eqnarray}
\frac{1}{a^{2}}\frac{\partial(a^{2}\vec{E})}{\partial\eta}-\nabla\times\vec{B}-\frac{n}{\eta^{2}}\frac{\vec{A}}{a^{2}}&=&4\pi(\vec{J}_{Y}+\vec{j})
\end{eqnarray}
in vector form where $\frac{n}{\eta^{2}}$ is equation (45). In equation (89) we have $\vec{J}_{Y}$ which is the current due to the flux of fluid $Y$ representing the $MD$ epoch (as a fluid) and $\vec{j}$ is the current due to plasma. Therefore, we have
\begin{eqnarray}
\vec{J}_{Y}+\vec{J}&=&\vec{J}_{T}
\end{eqnarray}
where $\vec{J}_{T}$ is the total current. Due to dominant contribution of plasma effects to that of flux current, $\vec{J}_{Y}$ or mathematically, when $|\vec{J_{T}}|=J_{T}=\sqrt{J^{2}_{Y}+J^{2}-2J_{Y}J\cos\theta}=J\sqrt{1+\frac{J^{2}_{Y}}{J^{2}}-\frac{2J_{Y}\cos\theta}{J}}\approx J\Bigg(1+\frac{J^{2}_{Y}}{2J^{2}}-\frac{J_{Y}\cos\theta}{J}+\dots\Bigg)$ ($J\gg J_{Y}\cos\theta$) to linear order (as the higher-order terms are very minute or negligible), and if
\begin{eqnarray}
|\vec{j}|^{2}\gg|\vec{J}_{Y}|^{2}
\end{eqnarray}
we have $J_{T}\approx J$ hence, $|\vec{J_{T}}|\approx|\vec{J}|$ implying that
\begin{eqnarray}
\vec{J}_{T}&\approx&\vec{j}.
\end{eqnarray}
For further clarification please refer to appendix $A$. This then implies that equation (89) can be written as
\begin{eqnarray}
\frac{1}{a^{2}}\frac{\partial(a^{2}\vec{E})}{\partial\eta}-\nabla\times\vec{B}-\frac{n}{\eta^{2}}\frac{\vec{A}}{a^{2}}&=&4\pi\vec{j}
\end{eqnarray} in vector form.

We can see that equation (93) above is the same as equation (49). Therefore, from this point onwards [from equation (49)], we can use the same analysis as we did during the subsection of the $RD$ epoch. We will obtain the same equations and hence, the same conclusions as in the analysis in the subsection of the $RD$ epoch except that for this epoch, inflation-produced, large-scale magnetic fields will cross the Hubble horizon for the first time at the beginning of this epoch of $MD$, evolve superadiabatically until they cross the horizon for a second time much later during this epoch of $MD$ and then go back to adiabatic decay until the present time. This implies that before the beginning of this epoch of $MD$, inflation-produced, large-scale magnetic fields evolved adiabatically starting at the inflationary epoch. We now consider large-scale superadiabatic amplification of magnetic fields in all epochs starting from the $RD$ epoch in the next subsection.

\subsection{Large-scale superadiabatic magnetic amplification}

Considering the arguments after equation (7) in \cite{Tim32} up to section 3.2 in \cite{Tim32}, large-scale (causally disconnected) magnetic fields evolve in line with the power-law equations (74) and (77) from the time they are well outside the Hubble radius for the first time at the beginning of the $RD$ or matter-radiation equality transition or $MD$ epoch until the time of their re-entry much later during the $MD$ epoch (or dust era). Equations (74) and (77) are no longer valid once the magnetic fields are back inside the horizon \cite{Tim32}. From then onwards, the ideal-$MHD$ limit applies, the magnetic flux remains conserved and the magnetic field decays adiabatically \cite{Tim32} ($i.e.$ , $B\propto a^{-2}$).

We will now consider the evolution of superhorizon-sized cosmological magnetic fields after inflation ($i.e.,$ from the beginning of $RD$ or matter-radiation equality or $MD$ epoch until their re-entry after crossing the horizon for a second time much later during the $MD$ epoch and then they go back to adiabatic decay until the present time). After evaluating the four integration constants on the $RHS$ of equations (74) and (77), the equations recast into the form below \cite{Tim32, Tim38},
\begin{eqnarray}
 B&=&[B_{0}-\eta_{0}(2a_{0}H_{0}B_{0}+B^{\prime}_{0})]\Bigg(\frac{a_{0}}{a}\Bigg)^{2}+\eta_{0}(2a_{0}H_{0}B_{0}+B^{\prime}_{0})\Bigg(\frac{a_{0}}{a}\Bigg)^{2}\Bigg(\frac{\eta}{\eta_{0}}\Bigg)
\end{eqnarray}
where the relation $H=\frac{a^{\prime}}{a^{2}}$ was used for the Hubble parameter and the subscript $k$ is dropped for convenience sake \cite{Tim32} and without loss of generality (The subscript $0$ represents a given initial time and please check the notes \cite{Tim403}). The above equation (94) monitors the linear evolution of superhorizon-sized magnetic fields on spatially flat $FLRW$ backgrounds \cite{Tim32}. $w=w(t)$ meaning that the barotropic index of the matter is not necessarily constant but it can vary with time. Hence, equation (94) applies continuously throughout the lifetime of the Universe [from the beginning of the epoch of $RD$ to much later during the $MD$ epoch or the beginning of the epoch of matter-radiation equality to much later during the $MD$ epoch or the beginning of the $MD$ epoch to much later during the $MD$ epoch (or dust era)], provided the cosmological expansion is entirely smooth and the matter can always be treated as a single barotropic medium \cite{Tim32}. Under this condition, equation (94) also monitors the magnetic evolution through the matter-radiation equality transition \cite{Tim32}.

$w$ is believed to maintain constant value during prolonged periods in the lifetime of the Universe \cite{Tim32}. For example, the matter-radiation equality transit (epoch) which is much more prolonged than the inflation or reheating epoch \cite{Tim46}. The cosmological scale factor and the conformal time are related by \cite{Tim32}
\begin{eqnarray}
 a&=&a_{0}\Bigg(\frac{\eta}{\eta_{0}}\Bigg)^{\frac{2}{1+3\omega}}
\end{eqnarray}
as long as $w$ remains invariant where $w\neq-\frac{1}{3}$ and the zero suffix indicates a given initial time. Using the above equation (95), it is easy to show that $H=\frac{a^{\prime}}{a^{2}}=\frac{2}{(1+3\omega)a\eta}$ and then recast equation (94) into \cite{Tim32}
\begin{eqnarray}
 B&=&-\Bigg[\Bigg(\frac{4}{1+3\omega}-1\Bigg)B_{0}+\eta_{0}B^{\prime}_{0}\Bigg]\Bigg(\frac{a_{0}}{a}\Bigg)^{2}+\Bigg(\frac{4B_{0}}{1+3\omega}+\eta_{0}B^{\prime}_{0}\Bigg)\Bigg(\frac{a_{0}}{a}\Bigg)^{\frac{3(1-\omega)}{2}}.
\end{eqnarray}
This equation monitors the linear evolution of superhorizon-sized magnetic fields on spatially flat $FLRW$ backgrounds filled with a single barotropic medium \cite{Tim32}. The difference with equation (94) is that here the barotropic index of the matter has been treated as a constant \cite{Tim32}. As a result, equation (96) does not apply constantly throughout the evolution of the Universe, but only to periods during which $w=$constant$\neq\frac{-1}{3}$ $e.g.$, the matter-radiation equality epoch and the $RD$ epoch \cite{Tim46} (since the matter-radiation equality transit is prolonged enough to be assumed to be an epoch which is logical as explained before above, $w$ can be taken to be constant during this transit implying that $w$ does not evolve during this transit period). This means that equation (96) is a special case of equation (94) \cite{Tim32}.
 
Well outside the Hubble radius from the beginning of or during either the $RD$ epoch or matter-radiation equality transition (epoch) or $MD$ epoch until much later during the $MD$ epoch, large-scale magnetic fields on spatially flat $FLRW$ backgrounds obey the above equations (94) and (96) because they contain modes with decay rates slower than the adiabatic \cite{Tim32}. These slowly-decaying magnetic modes depend on their associated coefficients \cite{Tim32}. When the coefficients are of roughly the same order of magnitude, the slowly decaying modes quickly take over and dictate the subsequent evolution of the magnetic field \cite{Tim32}. This means that the initial conditions during inflation, at the beginning of the post-inflationary evolution of magnetic fields, the nature of the transitions and the beginning and end of the reheating epoch or $RD$ epoch are very important if we are considering inflation-produced magnetic fields to be outside the horizon at the beginning of the $RD$ epoch or $MD$ epoch. The moment when inflation-produced magnetic fields cross over to the epoch of $MD$ from the epoch of matter-radiation equality transition, the initial conditions for the evolution of the magnetic fields during the epoch of $MD$ are also important. The conditions during inflation, at the beginning of the post-inflationary evolution of magnetic fields, the nature of the transitions, the beginning and the end of the $RD$ epoch would determine the initial conditions for the evolution of the magnetic fields during the matter-radiation equality epoch and they are very important (that is, for inflation-produced magnetic fields that are outside the Hubble horizon at the beginning of this transition epoch). We now analyze the role of the initial conditions in the following section.
 
\section{The functions of initial conditions}
 
The initial conditions are decided by the field's behaviour in the de Sitter phase and by the nature of the transitions (in some cases epoch for matter-radiation equality transition) to the reheating or $RD$ or $MD$ epochs and the moment when inflation-produced magnetic fields crossover into the matter-radiation equality transition epoch from the $RD$ epoch or from the equality epoch to the $MD$ epoch. We will discuss one typical and matching initial-conditions scenario [for a detailed discussion please check (\cite{Tim32} and \cite{Tim41})].
 
\subsection{Initial-conditions scenario}
 
In this scenario \cite{Tim32}, the background barotropic index, $w$ changes abruptly from $w^{-}_{*}$ before a transition to $w^{+}_{*}$ afterwards (the $*$-suffix marks the moment the Universe crosses from one epoch to the next \cite{Tim32}. Also, the $-$ and $+$ superscripts indicate the end of the era just before the transition and the beginning of the next respectively \cite{Tim32}. When we assume the matter-radiation equality transition to be an epoch, then the $*$-suffix marks the moments when inflation-produced magnetic fields crossover from the $RD$ epoch to matter-radiation equality transition epoch and from matter-radiation equality transition epoch to $MD$ epoch). Assuming that the matching spatial hypersurface is that of constant conformal time, then this would translate into a jump in the expansion rate of the background Universe, on either side of the transition. This implies a discontinuity in the extrinsic curvature of the matching hypersurface, which requires the presence of a thin shell there with finite energy-momentum tensor \cite{Tim32}. Now, assuming that the width of the shell is too small compared to the scales of interest, would imply that the aforementioned shell can be replaced by a spacelike hypersurface \cite{Tim404}. Then discontinuities of this nature can be used to avoid the details of early cosmological transitions \cite{Tim32}. In this scenario, $w$ undergoes an abrupt change from $w^{-}_{*}$ before the transition to $w^{+}_{*}$ afterwards (with $w^{+}_{*}\neq w^{-}_{*}$). Note that for certain instances in this scenario, we will assume the matter-radiation equality transit to be an epoch which is logical as explained before above.
 
The usual inflationary magnetogenesis scenarios demand that the magnetic field decays adiabatically throughout the de Sitter era ($i.e.$, $B\propto a^{-2}$) \cite{Tim32}. On the other hand, the usual non-conventional mechanisms of primordial magnetic generation amplify their magnetic fields superadiabatically during inflation ($i.e.$, $B\propto a^{-m}$ with $0\leq m<2$)\cite{Tim32}, \cite{Tim43}, \cite{Tim44}. Now let us assume that all along the de Sitter phase the magnetic field obeys the law \cite{Tim32}
\begin{eqnarray}
 B&=&B_{0}\Bigg(\frac{a_{0}}{a}\Bigg)^{m}=B_{0}\Bigg(\frac{\eta}{\eta_{0}}\Bigg)^{m}
\end{eqnarray}
where $0\leq m\leq2$ and the zero suffix indicates the beginning of the exponential expansion ($w=-1$). Differentiating equation (97) with respect to the conformal time gives $B^{\prime}=\frac{mB}{\eta}$, which implies that, \cite{Tim32}
\begin{eqnarray}
 \eta^{-}_{*}B^{\prime-}_{*}&=&mB^{-}_{*}
\end{eqnarray}
at the end of inflation proper. Remember that for our study, inflation-produced magnetic fields evolved adiabatically until the beginning of either the $RD$ epoch or matter-radiation equality transition epoch or $MD$ epoch from inflation implying that $m=2$. During reheating $w$ changes from $w^{-}_{*}=-1$ to $w^{+}_{*}=0$ \cite{Tim32}. Then throughout reheating inflation-produced, large-scale magnetic fields evolved as
\begin{eqnarray}
 B&=&B^{+}_{*}\Bigg(\frac{a^{+}_{*}}{a}\Bigg)^{2}
\end{eqnarray}
with $a\geq a^{+}_{*}$. Due to the arguments in the first paragraph of this subsection above, constraint equation (98) translates into \cite{Tim32}
\begin{eqnarray}
 \eta^{+}_{*}B^{\prime+}_{*}&=&-mB^{+}_{*}.
\end{eqnarray}
The above sets the initial conditions for the evolution of the magnetic fields where we are following equation (99). Therefore, magnetic fields drop as $B\propto a^{-2}$ and the magnetic fields decay adiabatically throughout reheating.

Let us consider magnetic fields evolution in the subsequent epoch of $RD$. Following equation (99) and keeping in mind that $a\propto\eta^{2}$ during reheating, we deduce that $B\propto\eta^{-4}$ throughout that period. Then,
\begin{eqnarray}
 \eta^{-}_{*}B^{\prime-}_{*}&=&-4B^{-}_{*}
\end{eqnarray}
just before the transition to the epoch of $RD$. During that time, $w$ changes from $w^{-}_{*}=0$ to $w^{+}_{*}=\frac{1}{3}$ and equation (96) reads \cite{Tim32,Tim200}
\begin{eqnarray}
 B&=&-(B^{+}_{*}+\eta^{+}_{*}B^{\prime+}_{*})\Bigg(\frac{a^{+}_{*}}{a}\Bigg)^{2}+(2B^{+}_{*}+\eta^{+}_{*}B^{\prime+}_{*})\Bigg(\frac{a^{+}_{*}}{a}\Bigg)
\end{eqnarray}
with $a\geq a^{+}_{*}$ (please check the notes \cite{Tim403}). Due to the arguments in the first paragraph of this subsection, constraint equation (101) recasts into
\begin{eqnarray}
 \eta^{+}_{*}B^{\prime+}_{*}&=&-4B^{+}_{*}
\end{eqnarray}
and sets the initial conditions for the magnetic evolution in the $RD$ epoch. Substituting the above into the right hand side ($RHS$) of equation (102) results in
\begin{eqnarray}
 B&=&3B^{+}_{*}\Bigg(\frac{a^{+}_{*}}{a}\Bigg)^{2}-2B^{+}_{*}\Bigg(\frac{a^{+}_{*}}{a}\Bigg)
\end{eqnarray}
where $a\geq a^{+}_{*}$. As a result, superhorizon-sized magnetic fields are superadiabatically amplified ($i.e.,$ $B\propto a^{-1}$) all along the $RD$ epoch until much later during the $MD$ epoch when they cross the horizon for a second time and they go back to adiabatic decay until the present time.

Due to the arguments in paragraph 1 of section (IV.2.), it would be logical to infer that $w=\frac{1}{9}$ and we would have
\begin{eqnarray}
 B&=&-(2B^{-}_{*}+\eta^{-}_{*}B^{\prime-}_{*})\Bigg(\frac{a^{-}_{*}}{a}\Bigg)^{2}+(3B^{-}_{*}+\eta^{-}_{*}B^{\prime-}_{*})\Bigg(\frac{a^{-}_{*}}{a}\Bigg)
\end{eqnarray}
where $a\geq a^{-}_{*}$. Prior to equilibrium of the matter-radiation equality epoch, we find $\eta^{-}_{*}B^{\prime-}_{*}=-B^{-}_{*}$ since $a\propto\eta$ when $w=\frac{1}{3}$ for the $RD$ epoch. We are following equation (104). This implies that inflation-produced, large-scale magnetic fields are superadiabatically amplified all along the $RD$ epoch. But before then, they evolved adiabatically. Inserting the condition above in equation (105) results in
\begin{eqnarray}
B&=&-B^{-}_{*}\Bigg(\frac{a^{-}_{*}}{a}\Bigg)^{2}+2B^{-}_{*}\Bigg(\frac{a^{-}_{*}}{a}\Bigg)
\end{eqnarray}
where $a\geq a^{-}_{*}$. As a result, superhorizon-sized magnetic fields are superadiabatically amplified ($i.e.,$ $B\propto a^{-1}$) all along the matter-radiation equality epoch (which shows smooth evolution of magnetic fields from the $RD$ epoch) until much later during the $MD$ epoch when they cross the horizon for a second time and they go back to adiabatic decay until the present time. When we inferred that $w=\frac{1}{6}$ for the matter-radiation equality epoch we found that inflation-produced, large-scale magnetic fields evolved as $B\propto a^{-\frac{5}{4}}$ during this epoch and this shows that evolution of magnetic fields was not smooth from the $RD$ epoch to the matter-radiation equality epoch. Nevertheless, the magnetic fields were superadiabatically amplified during the matter-radiation equality epoch until much later during the $MD$ epoch when they crossed the horizon for a second time and they went back to adiabatic decay until the present time.

Following equation (106) and keeping in mind that $a\propto\eta^{\frac{3}{2}}$ during the equality epoch we deduce that $B\propto\eta^{-\frac{3}{2}}$ throughout that period. This implies that inflation-produced, large-scale magnetic fields evolved adiabatically until the beginning of the epoch of $RD$. During the matter-radiation equality epoch, inflation-produced, large-scale magnetic fields evolved superadiabatically. Then,
\begin{eqnarray}
\eta^{+}_{*}B^{\prime+}_{*}&=&-\frac{3}{2}B^{+}_{*}
\end{eqnarray}
at the start of the $MD$ epoch. Equation (96) reads \cite{Tim32,Tim203}
\begin{eqnarray}
 B&=&-(3B^{+}_{*}+\eta^{+}_{*}B^{\prime+}_{*})\Bigg(\frac{a^{+}_{*}}{a}\Bigg)^{2}+(4B^{+}_{*}+\eta^{+}_{*}B^{\prime+}_{*})\Bigg(\frac{a^{+}_{*}}{a}\Bigg)^{\frac{3}{2}}
\end{eqnarray}
where $a\geq a^{+}_{*}$ (please check the notes \cite{Tim403}).

Inserting equation (107) in equation (108) results in
\begin{eqnarray}
B&=&-\frac{3}{2}\Bigg(\frac{a^{+}_{*}}{a}\Bigg)^{2}+\frac{5}{2}\Bigg(\frac{a^{+}_{*}}{a}\Bigg)^{\frac{3}{2}}
\end{eqnarray}
where $a\geq a^{+}_{*}$. As a result, superhorizon-sized magnetic fields are superadiabatically amplified ($i.e.,$ $B\propto a^{\frac{-3}{2}}$) during the dust era as well until much later during this era when magnetic fields cross the horizon for a second time and they go back to adiabatic decay until the present time.

We now consider evolution of magnetic fields during the $MD$ epoch again. For this case, inflation-produced, large-scale magnetic fields evolve adiabatically until the beginning of the $RD$ epoch. We use the approach in \cite{Tim32} where the matter-radiation equality period is a transit (and not an epoch). Due to the arguments in the first paragraph of this scenario, the constraint $\eta^{-}_{*}B^{\prime-}_{*}=-B^{-}_{*}$ recasts into
\begin{eqnarray}
\eta^{+}_{*}B^{\prime+}_{*}&=&-B^{+}_{*}
\end{eqnarray}
at the beginning of the $MD$ epoch. During the time of the matter-radiation equality, $w$ changes from $w^{-}_{*}=\frac{1}{3}$ to $w^{+}_{*}=0$ and equation (96) reads as equation (108). Substituting equation (110) in equation (108) results in
\begin{eqnarray}
B&=&-2B^{+}_{*}\Bigg(\frac{a^{+}_{*}}{a}\Bigg)^{2}+3B^{+}_{*}\Bigg(\frac{a^{+}_{*}}{a}\Bigg)^{\frac{3}{2}}
\end{eqnarray}
where $a\geq a^{+}_{*}$. As a result, superhorizon-sized magnetic fields are superadiabatically amplified ($i.e.,$ $B\propto a^{-\frac{3}{2}}$) during the dust era as well until much later during the epoch of $MD$ when they cross the horizon for a second time and they go back to adiabatic decay until the present time.

We now consider evolution of magnetic fields during the matter-radiation equality transit, assuming that the transit is an epoch. Here we assume that inflation-produced magnetic fields evolve adiabatically until the beginning of matter-radiation equality transit (epoch). Therefore, we don't start from equation (102), instead we continue from equation (103) and assume that the evolution of magnetic fields during the $RD$ epoch are monitored by equation (99). With equation (103) as the initial conditions, magnetic fields decay adiabatically all along the $RD$ epoch.

We find that $\eta^{-}_{*}B^{\prime-}_{*}=-2B^{-}_{*}$ prior to the equilibrium time, since $a\propto\eta$ when $w=\frac{1}{3}$. This sets the initial conditions for the magnetic evolution during the matter-radiation equality epoch. Now, substituting the initial condition above into the $RHS$ of equation (105) results in
\begin{eqnarray}
 B&=&B^{-}_{*}\Bigg(\frac{a^{-}_{*}}{a}\Bigg)
\end{eqnarray}
where $a\geq a^{-}_{*}$. Therefore, superhorizon-sized magnetic fields are superadiabatically amplified ($i.e.,$ $B\propto a^{-1}$) all along the transit of the matter-radiation equality (epoch) until much later during the $MD$ epoch when they cross the horizon for a second time and they go back to adiabatic decay until the present time.

We now consider the evolution of magnetic fields during the $MD$ epoch and we assume the matter-radiation equality transit to be an epoch. Then, inflation-produced, large-scale magnetic fields evolved adiabatically until the beginning of the epoch of matter-radiation equality transit. During the matter-radiation equality epoch, inflation-produced, large-scale magnetic fields evolved superadiabatically. Then we will have condition (107) at the start of the $MD$ epoch. Then plugging equation (107) in equation (108) results in equation (109). This implies that superhorizon-sized magnetic fields are superadiabatically amplified ($i.e.,$ $B\propto a^{-\frac{3}{2}}$) during the dust era as well until much later during this era when magnetic fields cross the horizon for a second time and they go back to adiabatic decay until the present time.

Now assuming adiabatic decay during the matter-radiation equality epoch, we will deduce that $B\propto\eta^{-3}$ throughout that period. This implies that inflation-produced, large scale magnetic fields evolved adiabatically until the beginning of the $MD$ epoch and here we assume the matter-radiation equality period to be an epoch. Then,
\begin{eqnarray}
\eta^{+}_{*}B^{\prime+}_{*}&=&-3B^{+}_{*}
\end{eqnarray}
at the beginning of the $MD$ epoch. Inserting equation (115) in equation (108) results in
\begin{eqnarray}
B&=&B^{+}_{*}\Bigg(\frac{a^{+}_{*}}{a}\Bigg)^{\frac{3}{2}}
\end{eqnarray}
where $a\geq a^{+}_{*}$. As a result, superhorizon-sized magnetic fields are superadiabatically amplified ($i.e.,$ $B\propto a^{\frac{-3}{2}}$) during the dust era as well again until much later during the epoch of $MD$ when they cross the horizon for a second time and they go back to adiabatic decay until the present time.

We now consider the evolution of magnetic fields during the $MD$ epoch and we assume the matter-radiation equality period to be a transit and not an epoch. Inflation-produced magnetic fields evolved adiabatically until the beginning of the $MD$ epoch. Due to the arguments in the first paragraph of this subsection, the constraint $\eta^{-}_{*}B^{\prime-}_{*}=-2B^{-}_{*}$ recasts into
\begin{eqnarray}
 \eta^{+}_{*}B^{\prime+}_{*}&=&-2B^{+}_{*}
\end{eqnarray}
at the beginning of the $MD$ epoch. Substituting equation (107) in equation (108) results in
\begin{eqnarray}
 B&=&-B^{+}_{*}\Bigg(\frac{a^{+}_{*}}{a}\Bigg)^{2}+2B^{+}_{*}\Bigg(\frac{a^{+}_{*}}{a}\Bigg)^{\frac{3}{2}}
\end{eqnarray}
where $a\geq a^{+}_{*}$. As a result, superhorizon-sized magnetic fields are superadiabatically amplified ($i.e.,$ $B\propto a^{-\frac{3}{2}}$) during the dust era as well though not all along up to the end of the dust era.

Sometime much later during the dust era or $MD$ epoch when $\eta$ increases so much, inflation-produced, large-scale magnetic fields exit the horizon because by then the products $\mathcal{q}\eta$ and $k\eta$ eventually become larger than unity (in our study, the products $\mathcal{q}\eta$ and $k\eta$ eventually become larger than unity much later during the $MD$ epoch or dust era and not before that). Note that the time of horizon entry is also determined by the scale of the magnetic mode in question which determines the overall superadiabatic amplification. This then implies that equations (74) and (77) will no longer be valid and hence, magnetic fields will exit the horizon and go back to adiabatic decay until the present time \cite{Tim32}.

In summary, we find that for the case where inflation-produced, large-scale magnetic fields evolved adiabatically until the beginning of the epoch of $RD$ or epoch of matter-radiation equality, we find that magnetic fields evolve as $B\propto a^{-1}$, $B\propto a^{-1}$ and $B\propto a^{-\frac{3}{2}}$ during the epochs of $RD$, matter-radiation equality and $MD$ or as $B\propto a^{-1}$ and $B\propto a^{-\frac{3}{2}}$ during the epochs of matter-radiation equality and $MD$ respectively. This shows that magnetic fields evolve smoothly from the epoch of $RD$ until the end of the matter-radiation equality transit (epoch). But when we infer $w_{eff}=\frac{1}{6}$ (which is logical given the arguments in the first paragraph of subsection ($IV.2.$)), we find that magnetic fields evolve as $B\propto a^{-\frac{5}{4}}$ during the matter-radiation equality epoch and this shows that evolution of magnetic fields is not smooth from the epoch of $RD$ until the end of the matter-radiation equality epoch. Nevertheless, magnetic fields are superadiabatically amplified during the matter-radiation equality epoch for both values of $w_{eff}=\frac{1}{9}$ and $w_{eff}=\frac{1}{6}$. 

For the case where inflation-produced, large-scale magnetic fields evolved adiabatically until the beginning of the $RD$ epoch and where we assume the matter-radiation equality to be a transit (and not an epoch), we find that magnetic fields evolve as $B\propto a^{-1}$ and $B\propto a^{-\frac{3}{2}}$ during the epochs of $RD$ and $MD$ respectively.

Assuming that the matter-radiation equality is an epoch or transit and that inflation-produced, large-scale magnetic fields evolved adiabatically until the beginning of the epoch of $MD$, we find that magnetic fields evolved superadiabatically as $B\propto a^{-\frac{3}{2}}$ during the epoch of $MD$ until much later during the same epoch when they cross the horizon for a second time and go back to adiabatic decay until the present time.

Anyway, for all cases, some time much later during the epoch of $MD$, the inflation-produced, large-scale magnetic fields cross the horizon for a second time and they go back to adiabatic decay until the present time.

\section{Summary of findings for the epochs of $RD$, matter-radiation equality and $MD$ }

After examining equations (94) and (96) we notice that the first of the two magnetic modes on the $RHS$ decay adiabatically \cite{Tim32}.
\begin{figure}[h!]
\centering
\includegraphics[scale=0.35]{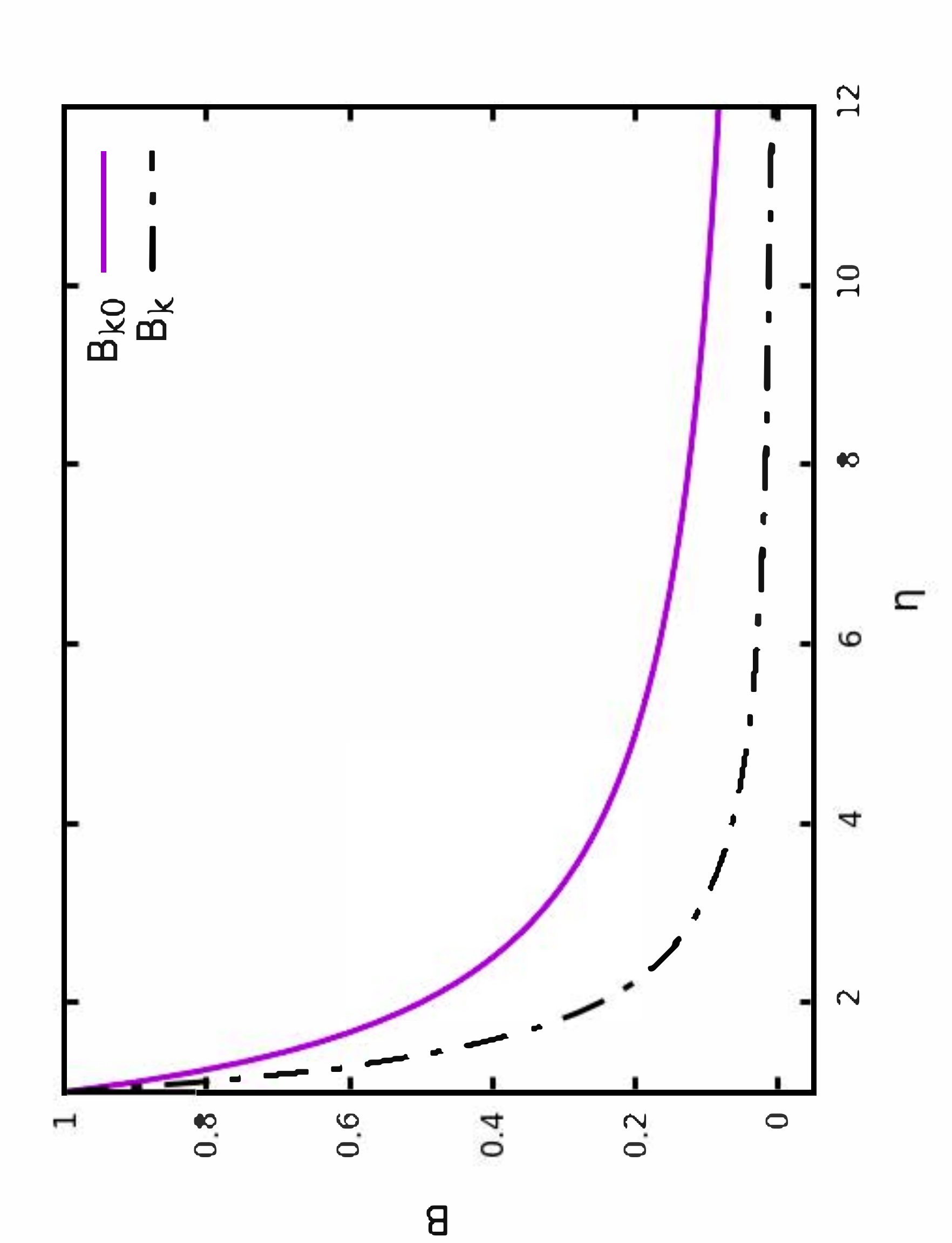}
\caption{This figure shows the typical behaviour of inflation-produced, large-scale magnetic fields from the beginning of or during the epochs of $RD$ or matter-radiation equality transit or $MD$ until much later during the $MD$ epoch when inflation-produced, large-scale magnetic fields cross the horizon for a second time and they go back to adiabatic decay until the present time. $B$ is the magnetic field while $\eta$ is the conformal time on the figure before above. The continuous line denoted by $B_{k0}$ is for superadiabatic amplification while the discontinuous line denoted by $B_{k}$ represents adiabatic magnetic decay. As explained before above, superadiabatic amplification may mean increase in strength of the magnetic field or slower magnetic decay rates than the standard (adiabatic) magnetic decay rate; in evolution of magnetic fields in cosmology, the latter is usually the case and the figure depicts that for superhorizon scales.}
\end{figure}
However, the rate of the second mode is not a priori fixed but depends on the $EoS$ of the cosmic medium \cite{Tim32}. The relation between the cosmological scale factor and the conformal time is determined by the latter. In particular, as long as $w$=constant$>$-$\frac{1}{3}$ the second mode on the $RHS$ of equation (96) decays at a rate slower than the adiabatic \cite{Tim32}. The same behaviour can also be seen in equation (94). Hence, when dealing with conventional matter, superhorizon-sized magnetic fields on spatially flat $FLRW$ backgrounds are superadiabatically amplified provided the initial conditions allow the second modes in equations (94) and (96) to survive and dominate \cite{Tim32}. The initial conditions of the post-inflationary magnetic evolution are determined by the field's behaviuor during inflation, at the beginning of the post-inflationary Universe, the nature of the transitions to the reheating epoch or the $RD$ epoch or the $MD$ epoch and the moment when inflation-produced, large-scale magnetic fields cross over into the transition (epoch) of matter-radiation equality from the epoch of $RD$ or from the equality epoch to the epoch of $MD$. Some of the initial-conditions scenarios are again discussed in detail in \cite{Tim32}.

Therefore, depending on the initial conditions, conventional large-scale magnetic fields can be superadiabatically amplified from the beginning of the epoch of $RD$ to much later during the epoch of $MD$ or from the beginning of the epoch of matter-radiation equality transit to much later during the epoch of $MD$ or from the beginning of the epoch of $MD$ to much later during the $MD$ epoch or dust era when inflation-produced, large-scale magnetic fields cross the horizon for a second time and they go back to adiabatic decay until the present time. Once inside the horizon, they decay adiabatically leading to magnitudes much higher than $10^{-53}G$ which would be the magnitude at present time due to adiabatic magnetic decay throughout the post-inflationary evolution of the Universe until the present time. This shows that by appealing to causality, the generalised cosmological Ohm's law and both standard and modified Maxwell's equations, one can increase the final strength of conventional inflationary large-scale magnetic fields by more orders of magnitude.

Similar equations to equations (72) and (75) were found in \cite{Tim39} and it was found (after using equations (72) and (75)) in the mentioned article that conventional inflationary, large-scale magnetic fields are superadiabatically amplified during certain periods of time of the post-inflationary Universe.

\section{Discussions and conclusions}

Adiabatic decay of magnetic fields on flat $FLRW$ spacetime translates into magnetic strengths below $10^{-50} G$ at present time \cite{Tim400}. As far as we know currently, such fields can never seed the galactic dynamo as mentioned before earlier or can never affect the dynamics of our Universe. Therefore, our goal in this paper is to show that superadiabatic amplification is possible using the generalised cosmological Ohm's law and both standard and modified Maxwell's equations on flat $FLRW$ spacetime which translates into magnetic strengths above $10^{-50}$ in the present time and hence, seed the galactic dynamo or affect the dynamics of our Universe.

The consequences of the modified Maxwell's equations were investigated in \cite{Tim27}. For reasonable parameters it was shown that modification does not affect existing experiments and observations \cite{Tim27}. Nevertheless, it is argued that, the field equations coupled with a curvature term can be testable in astrophysical environments where the mass density is high or the gravity of electromagnetic radiation plays a dominant role in the dynamics of the Universe, $e.g.,$ the interior of neutron stars and the early Universe \cite{Tim27}.

We have examined ($i$) the single-fluid and multi-fluid approximations and used them to derive the modified Maxwell tensors, ($ii$) the generalised cosmological Ohm's law valid on both cosmological horizon (and superhorizon) scales and used it in both standard and modified electromagnetism, ($iii$) the terms $RA^{2}$ , $R_{ab}A^{a}A^{b}$ and $RA^{2}+R_{ab}A^{a}A^{b}$ and their significance in the evolution of cosmological magnetic fields in the post-inflationary Universe, ($iv$) the post-inflationary evolution of cosmological magnetic fields using the generalised cosmological Ohm's law, the flux currents of both radiation and matter fluids and the modified Maxwell tensors derived from both single-fluid and multi-fluid formalisms. The evolution of the inflation-produced, large-scale magnetic fields are examined from the beginning of or during the epoch of $RD$ , matter-radiation equality and the $MD$ until much later during the $MD$ epoch with initial conditions starting during the inflationary era until the beginning of the $MD$ epoch. We find that cosmological magnetic fields are superadiabatically amplified from the beginning of the epoch of $RD$ to much later during the $MD$ epoch or from the beginning of the epoch of matter-radiation equality transition to much later during the $MD$ epoch or from the beginning of $MD$ epoch itself to much later during the epoch of $MD$ when the magnetic fields exit the horizon for a second time and they go back to adiabatic decay until the present time and this is possible after using the generalised cosmological Ohm's law and both standard and modified Maxwell equations.

From the beginning of the epoch of $RD$ to much later during the $MD$ epoch, the magnetic-decay rate slows down as $B\propto a^{-1}$ during the $RD$, $B\propto a^{-1}$ or $B\propto a^{-\frac{5}{4}}$ during the matter-radiation equality transit (epoch), $B\propto a^{-\frac{3}{2}}$ during the $MD$ epoch until its re-entry inside the horizon much later during the $MD$ epoch and due to the reasons in subsection 4.1 of \cite{Tim32}, the magnetic-decay rate will be adiabatic until the present time. From the beginning of the epoch of the matter-radiation equality transition to much later during the $MD$ epoch, the magnetic-decay rate slows down as $B\propto a^{-1}$ or $B\propto a^{-\frac{5}{4}}$ during the matter-radiation equality transit (epoch), $B\propto a^{-\frac{3}{2}}$ during the $MD$ epoch until its re-entry inside the horizon much later during the $MD$ epoch and due to the reasons mentioned before above, the magnetic-decay rate will be adiabatic until the present time. From the beginning of the $MD$ epoch to much later during the $MD$ epoch, the magnetic-decay rate slows down as $B\propto a^{-\frac{3}{2}}$ during the $MD$ epoch until its re-entry inside the horizon much later during the $MD$ epoch and due to the reasons mentioned before above, the magnetic-decay rate will be adiabatic until the present time. This confirms that superadiabatic amplification of large-scale magnetic fields generated during inflation by any mechanism imaginable is possible in flat $FLRW$ spacetime. This is true for both cases where we used the modified Maxwell's equations, fluid flux currents and the generalised cosmological Ohm's law and the case where we used standard Maxwell's equations, fluid flux currents and again the generalised cosmological Ohm's law.  We have also shown that the single-fluid and multi-fluid formalisms are resourceful tools for investigating the evolution of cosmological magnetic fields throughout the post-inflationary Universe especially during the relatively explored epoch of the matter-radiation equality transit where we employed the multi-fluid formalism. Due to the reasons mentioned before earlier in the first paragraph of section ($IV$.2.) and given that the so-called transit of matter-radiation equality is much more prolonged than the inflationary or reheating epochs, the evolutionary history of inflation-produced, large-scale magnetic fields had to be examined during this transit (epoch). We assumed that $w$ is constant during the matter-radiation equality transit (epoch) which is logical as mentioned before earlier. Examining the evolution of magnetic fields during this so-called transit (epoch) of matter-radiation equality enriches the standard picture of the evolutionary history of inflation-produced, large-scale magnetic fields as it is much more prolonged than the inflationary and reheating epochs. As a result, magnetic fields would change much more appreciably during this quite long epoch (or so-called transit) than during the inflationary and reheating epochs. Therefore, with all this we see why magnetic fields with strengths of $10^{-6}G$ were detected.

\section{Acknowledgments}

This study was supported through the $UCT$ postgraduate funding office, $URC$ and $NGP$, $UCT$, Cape Town, South Africa.

\section{Data availability statement}

The data that supports the findings of this study are available within the article.

\section{Appendix A}

Plasma effects have dominant contributions to either of the flux currents, $\vec{J}_{X}$ or $\vec{J}_{Y}$ or the total of the flux currents, $\vec{J}_{X}+\vec{J}_{Y}$. From the equations between equations (85) and (86) before above, for $-1\leq\cos\theta\leq1$, we are given that
\begin{eqnarray}
J^{2}_{Z}&=&J^{2}_{X}+J^{2}_{Y}-2J_{X}J_{Y}\cos\theta .
\end{eqnarray}
For $\cos\theta=-1$, equation (117) will translate into
\begin{eqnarray}
J^{2}_{Z}&=&J^{2}_{X}+2J_{X}J_{Y}+J^{2}_{Y} \nonumber\\
         &=&(J_{X}+J_{Y})^{2} .
\end{eqnarray}
Then,
\begin{eqnarray}
\frac{J^{2}_{Z}}{J^{2}}&=&\Bigg(\frac{J_{X}+J_{Y}}{J}\Bigg)^{2} \nonumber\\
                       &=&\Bigg(\frac{J_{X}}{J}+\frac{J_{Y}}{J}\Bigg).
\end{eqnarray}
When $J\gg J_{X}$ and $J\gg J_{Y}$, then $\frac{J^{2}_{Z}}{J^{2}}\sim0$ implying that $\frac{J_{Z}}{J}\sim0$.

 For $\cos\theta=1$, equation (117) translates into
\begin{eqnarray}
 J^{2}_{Z}&=&J^{2}_{X}-2J_{X}J_{Y}+J^{2}_{Y} \nonumber\\
          &=&(J_{X}-J_{Y})^{2} \nonumber\\
          &=&(J_{Y}-J_{X})^{2}.
\end{eqnarray}
Then,
\begin{eqnarray}
 \frac{J^{2}_{Z}}{J^{2}}&=&\Bigg(\frac{J_{X}-J_{Y}}{J}\Bigg)^{2} \nonumber\\
                        &=&\Bigg(\frac{J_{X}}{J}-\frac{J_{Y}}{J}\Bigg)^{2} \nonumber\\
                        &=&\Bigg(\frac{J_{Y}}{J}-\frac{J_{X}}{J}\Bigg)^{2}.
\end{eqnarray}
When $J\gg J_{X}$ and $J\gg J_{Y}$, then $\frac{J^{2}_{Z}}{J^{2}}\sim0$ implying that $\frac{J_{Z}}{J}\sim0$. Then for $-1\leq\cos\theta\leq1$ $\frac{J^{2}_{Z}}{J^{2}}\sim0$ or $\frac{J_{Z}}{J}\sim0$. Note that $J\gg J_{X}|\cos\theta|$ as $J\gg J_{X}$. Similarly, $J\gg J_{Y}|\cos\theta|$ as $J\gg J_{Y}$.


\begin{thebibliography}{99}
\bibitem{Tim100} Note that we still use $\tilde{B}^{\prime\prime}_{k}$ and $\tilde{B}_{k}$ instead of $\tilde{B}^{\prime\prime}_{k1}$ and $\tilde{B}_{k1}$ (hence, differentiating standard electromagnetism from modified electromagnetism) for standard electromagnetism because the behavior of magnetic fields in both modified and standard electromagnetism is the same as we will find out later on in this study.
\bibitem{Tim403} Note that to obtain equations (73) and (76) we used the generalised cosmological Ohm's law at the beginning of the epoch of $RD$ or matter-radiation equality or $MD$ on horizon scales and arguments in \cite{Tim32} in a different context from that in the just mentioned article. In reference \cite{Tim32} again, equation (76) was obtained in the absence of currents and hence, represents magnetic fields as they cross the horizon during the de Sitter era in the epoch of inflation. Nevertheless, we obtain the same equations (77), (94), (96), (102), and (108) in this study as in references \cite{Tim32,Tim200,Tim203}.
\bibitem{Tim37} The electrical properties of a medium are reflected in the simple Ohm's law throughout the post-inflationary Universe on subhorizon scales \cite{Tim32}. As a consequence, to eliminate a superhorizon-sized electric field, requires the presence of currents coherent on the same scales \cite{Tim32}. Given that causality can never affect superhorizon-sized perturbations, the electric field will not vanish, unless it is fragmented into individually causally connected parts \cite{Tim32}. Nevertheless, even if we assume that the electric field has somehow been fragmented and eliminated by the local currents within the Hubble horizon, its superhorizon-sized magnetic counterpart is unaware so to speak of that for as long as it remains causally disconnected \cite{Tim32}.
\bibitem{Tim38} After calculating the integration constants of equations (74) and (77) we have $C_{1}=C_{3}=[B_{0}-\eta_{0}(2a_{0}H_{0}B_{0}+B^{\prime}_{0})]a^{2}_{0}$ and $C_{2}=\frac{\eta_{0}(2a_{0}H_{0}B_{0}+B^{\prime}_{0})a^{2}_{0}}{\mathcal{q}\eta_{0}}$ and $C_{4}=\frac{\eta_{0}(2a_{0}H_{0}B_{0}+B^{\prime}_{0})a^{2}_{0}}{k\eta_{0}}$ \cite{Tim32}. Now, since we have that $\mathcal{q}\eta_{0}\ll1$ and $k\eta_{0}\ll1$ on super-Hubble scales, it can be shown that $C_{2}\gg C_{1}$ and $C_{4}\gg C_{3}$ (unless $2a_{0}H_{0}B_{0}+B^{\prime}_{0}=0$ ) \cite{Tim32}. This explains why the second mode of equations (74) and (77) should not be discarded before evaluating the integration constants first \cite{Tim32}.
\bibitem{Tim404} Note that we can set $\eta^{+}_{*}=-\eta^{-}_{*}$ where $\eta^{+}_{*}>0$ and $\eta^{-}_{*}<0$ for the cosmological transition leading to the reheating epoch and set $\eta^{+}_{*}=\eta^{-}_{*}$ for the cosmological transitions leading to the epochs of $RD$ and $MD$. Consequently, assuming that there is no magnetic discontinuity at the linear level or that the magnetic evolution through a transit is smooth implies that $B^{+}_{*}=B^{-}_{*}$ and $B^{\prime+}_{*}=B^{\prime-}_{*}$ \cite{Tim32}.
\end{thebibliography}
\end{document}